\documentclass[letterpaper,12pt]{article}
\usepackage{amsmath}
\usepackage{graphicx}
\usepackage{enumerate}
\usepackage{natbib}
\usepackage{url} 

\newcommand{\blind}{1}

\usepackage{booktabs}
\usepackage{float}
\usepackage{setspace} 
\usepackage{algpseudocode}
\usepackage{algorithm}
\usepackage{tikz}
\usetikzlibrary{decorations.pathreplacing,calc}
\newcommand{\tikzmark}[1]{\tikz[overlay,remember picture] \node (#1) {};}
\newcommand*{\SpaceReservedForComments}{.5cm}%
\newcommand*{\HorizontalOffset}{-.55em}%
\newcommand*{\VerticalOffset}{0.7ex}%
\newcommand*{\AddNote}[4][]{%
    \begin{tikzpicture}[overlay, remember picture]
        \draw [decoration={brace,amplitude=0.3em},decorate,ultra thick,red, #1]
            ($(#3)+(\HorizontalOffset,-\VerticalOffset)$) --  ($(#2)+(\HorizontalOffset,\VerticalOffset)$)
            node [align=left, text width=\SpaceReservedForComments-0em, pos=0.5, anchor=east] {#4};
    \end{tikzpicture}
}%

\makeatletter
    \algrenewcommand\alglinenumber[1]{\tikzmark{\arabic{ALG@line}}\tiny#1:}
\makeatother

\newcommand{\bb}[1]{\boldsymbol{#1}} 
\newcommand{\bftab}{\fontseries{b}\selectfont}
\newcommand{\indep}{\perp \!\!\! \perp}
\begin{document}

\def\spacingset#1{\renewcommand{\baselinestretch}%
{#1}\small\normalsize} \spacingset{1}


\if1\blind
{
  \title{\bf Exploratory Hidden Markov Factor Models for Longitudinal Mobile Health Data: Application to Adverse Posttraumatic Neuropsychiatric Sequelae}
    \author{Lin Ge, Xinming An, Donglin Zeng, \\Samuel McLean, Ronald Kessler, and Rui Song\footnote{Lin Ge is graduate student (E\--mail: lge@ncsu.edu) and Rui Song is Professor (rsong@ncsu.edu), Department of Statistics, North Carolina State University, Raleigh, NC 27695. Xinming An (E-mail: Xinming\_An@med.unc.edu) is Research Assistant Professor, Department of Anesthesiology, The University of North Carolina at Chapel Hill, Chapel Hill, NC 27514. Donglin Zeng is Professor (E-mail: dzeng@bios.unc.edu), Department of Biostatistics, The University of North Carolina at Chapel Hill, Chapel Hill, NC 27599. Samuel McLean is Professor (E-mail: samuel\_mclean@med.unc.edu), Department of Psychiatry, The University of North Carolina at Chapel Hill, Chapel Hill, NC 27514. Ronald Kessler is Professor (E-mail: kessler@hcp.med.harvard.edu), Department of Health Care Policy, Harvard Medical School, Boston, MA 02115. The research is partially supported by NSF under Grant DMS-1555244 and 2113637, NIMH under Grant U01MH110925, the US Army Medical Research and Material Command, The One Mind Foundation, and The Mayday Fund.}}
    \date{}
  \maketitle
} \fi

\if0\blind
{
  \bigskip
  \bigskip
  \bigskip
  \begin{center}
    {\LARGE\bf Exploratory Hidden Markov Factor Models for Longitudinal Mobile Health Data: Application to Adverse Posttraumatic Neuropsychiatric Sequelae}
\end{center}
  \medskip
} \fi

\bigskip
\begin{abstract}
\noindent Adverse posttraumatic neuropsychiatric sequelae (APNS) are common among veterans and millions of Americans after traumatic exposures, resulting in substantial burdens for trauma survivors and society. Despite numerous studies conducted on APNS over the past decades, there has been limited progress in understanding the underlying neurobiological mechanisms due to several unique challenges. One of these challenges is the reliance on subjective self-report measures to assess APNS, which can easily result in measurement errors and biases (e.g., recall bias). To mitigate this issue, in this paper, we investigate the potential of leveraging the objective longitudinal mobile device data to identify homogeneous APNS states and study the dynamic transitions and potential risk factors of APNS after trauma exposure. To handle specific challenges posed by longitudinal mobile device data, we developed exploratory hidden Markov factor models and designed a Stabilized Expectation-Maximization algorithm for parameter estimation. Simulation studies were conducted to evaluate the performance of parameter estimation and model selection. Finally, to demonstrate the practical utility of the method, we applied it to mobile device data collected from the Advancing Understanding of RecOvery afteR traumA (AURORA) study. 
\end{abstract}

\noindent%
\vfill

\newpage
\spacingset{1.9} 

\section{Introduction}\label{Intro}
Adverse posttraumatic neuropsychiatric sequelae (APNS) (e.g., pain, depression, and PTSD) are frequently observed in civilians and military veterans who have experienced traumatic events, such as car accidents and sexual assault. These APNS increase the risk of chronic illnesses, including cancer and heart disease, and substantially contribute to drug abuse, suicide, and disability. Moreover, APNS impose enduring psychosocial and financial burdens not only on individuals with the disorder but also on their families, communities, and society as a whole. 

However, little progress has been made in advancing APNS research over the past few decades due to several unique challenges. First, APNS have been evaluated through subjective self-reported measures, which lack objective reliability. Second, the heterogeneity among patients, as recognized in traditional classification and diagnoses, complicates the study of APNS. Lastly, these APNS disorders are often studied and treated independently, despite their frequent co-occurrence \citep{mclean2020aurora}. These obstacles hinder the identification of objective markers, the advancement in understanding the neurobiological mechanisms of APNS, and the development of effective preventative/treatment strategies. 

Identifying homogeneous states and exploring the dynamic prognosis of APNS in the immediate aftermath of trauma exposure holds promise for enhancing our understanding of APNS and identifying effective intervention options and appropriate timing at the individual level. Regrettably, due to the lack of appropriate data and effective statistical method, no large-scale studies have been conducted to investigate the onset, dynamic transitions (such as recovery and relapse), and associated risk factors of APNS. 
To help address the challenges, the National Institutes of Mental Health, joined by the US Army Medical Research and Material Command, several foundations, and corporate partners, developed the Advancing Understanding of RecOvery afteR traumA (AURORA) study \citep{mclean2020aurora}. This study gathered extensive biobehavioral data from a large cohort of trauma survivors (n = 2,997) across the United States, including self-reported surveys, web-based neurocognitive tests, digital phenotyping data (i.e., wrist wearable and smartphone data), psychophysical tests, neuroimaging assessments, and genomics data. Data collection starts in the early aftermath of the traumatic event and continues for a year. 

Leveraging this rich dataset, our work aims to mitigate the difficulties associated with APNS by i) identifying homogeneous latent states, ii) studying dynamic transition patterns over time, and iii) investigating potential risk factors of state transition. In contrast to previous studies that attempted to identify homogeneous subgroups for APNS relying on self-report survey data or neuroimaging data \citep{marquand2016beyond}, we focus on utilizing objective mobile device data, which tracks an individual's behavior, mood, and health in real-time, real-life environments. 
To achieve these goals, we develop both discrete-time and continuous-time exploratory hidden Markov factor models that can simultaneously identify homogeneous subtypes, investigate subtype-specific structure, and model individuals' progression and associated risk factors based on multivariate longitudinal data. 

Hidden Markov Models (HMMs) \citep{baum_statistical_1966} have been widely used in various fields \citep{mor2021systematic}. However, mobile device data presents two unique challenges that standard HMMs cannot handle, including the interdependent variables with unknown interrelationship structures and unevenly spaced measurements.

Mobile device sensor data, such as accelerometer data and photoplethysmography (PPG) from smartwatches, are highly intensive time series data. Typically, these raw data are preprocessed, and features are extracted using data processing pipelines within a larger time window (e.g., daily activity features derived from accelerometer data). These features are often technical summaries representing different characteristics of each time series variable and hence are often highly correlated. Due to the exponentially growing number of parameters in the covariance matrix as the number of features increases, assuming a fully free covariance matrix is infeasible. Therefore, under the HMM framework, features are commonly assumed to be independent, given the latent state membership. However, this assumption is often violated in real-world applications. 

To model the association between features, Factor analysis models (FMs) \citep{kim1978factor} provide an efficient and parsimonious approach and have been incorporated into HMMs in various ways. For example, the factor analyzed hidden Markov model \citep{rosti2002factor} combines an FM with a discrete-time HMM (DTHMM) \citep{vermunt1999discrete}, and has been extensively used in a variety of real-world applications, including speech recognition \citep{rosti2004factor}, environmental protection \citep{maruotti2017dynamic}, and seizure detection \citep{madadi2019hidden}. Similarly, \citet{Liu2016} introduced the regime-switching factor model to handle high-dimensional financial market data. However, they all assume homogeneous transition probability matrices, limiting their ability to account for the heterogeneity of transition probabilities over time and among different subjects and explore risk factors of state transition. 
To simultaneously capture the interrelationships among observed features and account for the variability of transition probabilities, a joint framework incorporating HMM, FM, and a feature-based transition model was recently proposed \citep{Song2017,zhou2022joint}. However, it is not directly applicable to mobile device data. Firstly, the framework employs a confirmatory factor model (CFM) with pre-specified structures for the factor loading matrices, which is not suitable for mobile device data that lacks such prior knowledge. Therefore, an exploratory factor model (EFM) is needed to explore the interrelationship structure among all observed features. Secondly, the framework assumes ordered states using the continuation-ratio logit model, which is inappropriate for analyzing AURORA data.

Another challenge posed by mobile device data is the irregular spacing of measurements. For example, activity and heart rate variability (HRV) data were collected only when the participants wore the watches, resulting in non-uniformly spaced observations and significant variation in sampling schedules between individuals. While the aforementioned methods are all based on DTHMM, assuming evenly spaced measurements and neglecting the impact of time gaps between consecutive observations on transition rates, continuous-time discrete-state HMM (CTHMM) was developed to handle irregularly spaced measurements \citep{cox2017theory}. CTHMM and its extensions that incorporate covariates to characterize transition rates are widely used in medical research that typically involves irregularly collected clinical measures \citep{liu2015efficient,lange2018estimating, amoros2019continuous,zhou2020continuous}. However, none of them focus on the interrelationships among features.

In this paper, to simultaneously address the two challenges and examine heterogeneous transition patterns, we propose an innovative model consisting of three components, including DTHMM/CTHMM, EFM, and multinomial logistic/log-linear transition model. 
Our contributions can be summarized as follows. 
First, we examine the utility of data collected in an open environment from consumer-grade mobile devices for mental health research. This contrasts with most existing studies on data collected in controlled lab environments. Second, we propose two Exploratory Hidden Markov Factor Models (EHMFM) that address the unique challenges introduced by mobile device data and depict non-homogeneous state transition processes of multiple individuals. While the Discrete-Time EHMFM assumes consistent time intervals, the Continuous-Time EHMFM accepts different structures of longitudinal data collected on a regular or irregular basis. 
Simulation studies using synthetic data demonstrate exceptional parameter estimation and model selection performance. Finally, we analyze HRV and activity data from the AURORA study, followed by interpretations and discussions of biological findings that highlight the immense potential of mobile health data and our proposed method for mental health research.


\section{AURORA Dataset}\label{Descrip Data}
This study focuses on two subsets of the AURORA data, each representing a distinct data structure of research interest. The first subset includes observations systematically collected every ten days from 180 patients. The second subset consists of irregularly sampled observations from 258 patients, with each patient providing at least 50 observations. Both subsets include 23 features, with four derived from activity data and the remaining 19 derived from  HRV data. See Appendix A for a detailed description of the variables and our data preprocessing approach. 

\section{Exploratory Hidden Markov Factor Model (EHMFM)} \label{model}

Motivated by the structures of the processed AURORA datasets, we consider data in the form of repeated measurements of $p$ features over $T_{i}$ occasions for each individual $i$ of $N$ subjects. The proposed models are in the framework of HMM. 
Let $w_{it}$ be the latent state of individual $i$ at occasion $t$, taking value from the finite discrete set $\{1,\cdots,\emph{J}\}$. Here, \emph{J} is the total number of states, which is fixed and known. Let $\bb{W}_{i}=(w_{i1},\cdots,w_{iT_{i}})$ be the state sequence over $T_{i}$ repeated measurements. Let a $\mit{J}\times \mit{J}$ matrix $\mit{\bb{P}_{it}}$ be the transition probability matrix for individual $i$ at occasion $t$, $t=\{2,\cdots,T_{i}\}$, of which the $(k,j)$ entry is $\bb{P}_{it,kj}=P(w_{it}=j|w_{i,t-1}=k)$, and $\bb{P}_{it,kk}=1-\sum_{j:j\neq k}\bb{P}_{it,kj}$. At $t=1$, we assume that the initial state follows a multinomial distribution with probabilities $\bb{\pi}=(\pi_{1},\cdots,\pi_{J})^{'}$, such that $\sum_{i=1}^{J}\pi_{i}=1$. The objective of the HMM is to delineate latent Markov processes given observations by estimating the transition probability matrix $\bb{P}$ and the initial state distribution.

Unlike the conventional HMM, our model incorporates two additional components to address the unique challenges and achieve our goals. In the first component, discussed in Section \ref{Factor model}, we posit a state-specific measurement model for the observations to learn interrelationship structures via EFM. In the second component, discussed in Section \ref{transition model}, we introduce transition models (TM) for learning heterogeneous transition patterns. 



\subsection{State-Specific Measurement Model} \label{Factor model}
Let $\mit{\bb{y}_{it}}$ denote a $\mathit{p} \times 1$ vector of the observed value of $p$ outcome variables for subject $i$ at time $t$. 
$\mit{\bb{z}_{it}}$ is a $K$ dimensional vector of latent scores assumed to be independent of $w_{it}$ and following a standard multivariate normal distribution. While we assume that $K$ is constant across states, our model can easily be extended to accommodate varying $K_{j}$. For each individual $i$, $\mit{\bb{Y}_{i}}=(\mit{\bb{y}_{i1}},\cdots,\mit{\bb{y}_{iT}})$ is a $p \times T$ matrix containing all measurements
and $\mit{\bb{Z}_{i}}=(\mit{\bb{z}_{i1}},\cdots,\mit{\bb{z}_{iT}}$) is a $K\times T$ matrix containing all latent features.
 
The first component of our model is an FM, with the primary goal of identifying the interrelationship structures between observed response variables and the underlying constructions of latent variables. For individual $i$ at time $t$, given $w_{it}=j$, the FM assumes the following state-specific measurement model:
\begin{equation}[\bb{y}_{it}|w_{it}=j]=\bb{\mu}_{j}+\bb{\Lambda}_{j}\bb{z}_{it}+\bb{e}_{it};\quad
     \bb{z}_{it}\stackrel{i.i.d.}{\sim}\mathcal{N}(\bb{0},\bb{I}_K),
     \bb{e}_{it}\stackrel{i.i.d.}{\sim}\mathcal{N}(\bb{0},\bb{\Psi}), \bb{z}_{it}\indep \bb{e}_{it} \label{cond dist yz}
\end{equation}
where $\bb{\mu}_{j}$ is a $p\times 1$ vector of state-specific expected mean response, $\bb{\Lambda}_{j}$ is a $p \times K$ state-specific factor loading matrix, $\bb{\Psi}$ is a $p \times p$ diagonal covariance matrix for the error term $\bb{e}_{it}$ with positive nonconstant diagonal entries. 
Alternatively, the model (\ref{cond dist yz}) can be expressed as $[\bb{y}_{it}|w_{it}=j]\stackrel{i.i.d.}{\sim}\mathcal{N}(\bb{\mu}_{j},\bb{\Lambda}_{j}\bb{\Lambda}_{j}^{'}+\bb{\Psi})$. It is crucial to emphasize that, unlike CFM with pre-specified structures of factor loading matrices, our approach does not impose any assumptions on $\bb{\Lambda}_j$. Therefore, the structure of $\bb{\Lambda}_j$ will be completely data-driven, making the first component of our model (\ref{cond dist yz}) an EFM.

\subsection{Transition Model} \label{transition model}
Considering the two data structures discussed in Section \ref{Descrip Data}, the appropriate transition models vary based on the data at hand. To provide a basic understanding of the structure of the proposed integrated model, we first illustrate the transition model in the DT-EHMFM in Section \ref{DT-EHMFM}, which ignores the effects of time intervals between two consecutive observations. In other words, the DT-EHMFM assumes consistent time intervals between measurements, which are frequently violated in mobile health data. Therefore, we subsequently introduce the CT-EHMFM in Section \ref{CT-EHMFM}, which relaxes the assumption of identical time intervals to allow longitudinal data to be collected irregularly.

\subsubsection{DT-EHMFM}\label{DT-EHMFM}
Given a state sequence $\bb{W}_{i}$, standard assumptions of DTHMM assume that 1) given a state $w_{it}$, observations $\bb{y}_{it}$ are independent, and 2) given a state $w_{it}$ and subjects' contextual features, the state at the subsequent occasion $w_{i,t+1}$ is unrelated to any information from previous occasions. 
Utilizing subjects' contextual information, we use a multinomial logistic regression model to explicitly characterize the transition probability matrix $\mit{\bb{P}_{it}}$ as follows:
\begin{equation}
    \log(\frac{\bb{P}_{it,kj}|\bb{x}_{it}}{\bb{P}_{it,kk}|\bb{x}_{it}})=\bb{x}_{it}^{'}\bb{B}_{kj},\textit{ }t=\{2,\cdots,T_{i}\},
\end{equation}\label{CTHMM}where $\bb{x}_{it}$ is a $d\times 1$ vector of vector of covariates for individual $i$ at time $t$, and $\bb{B}_{kj}$ is a state-specific $d\times 1$ vector of fixed effects coefficients.  
Here, $\bb{x}_{it}$ can be reduced to $\bb{x}_{i}$, which contains only baseline features. 
The $\bb{B}_{kj}$ intends to quantify the effect of covariates on the probability of transitioning from state $k$ to a different state $j$ to provide an understanding of how covariates influence transition patterns and investigate the potential risk factors. Conventionally, $\bb{B}_{kk}=\bb{0}$.

However, extreme caution is required when interpreting the predicted transition model in the case of direct application of DT-EHMFM to irregularly spaced datasets. When data are sampled irregularly, multiple complex transitions can occur between any two consecutive observations. Intuitively, given $w_{it}$, the distribution of $w_{i,t+1}$ tends to approach a uniform distribution as the interval between observations lengthen. Therefore, additional bias will be introduced when directly applying the DT-EHMFM on a dataset with varying time intervals. 

\subsubsection{CT-EHMFM}\label{CT-EHMFM}
Contrary to the DTHMM, which requires equal time intervals, the CTHMM takes into account the effects of the time interval. Thus, instead of directly depending on the transition probability matrix $\bb{P}$, the continuous-time Markov process is characterized by a transition intensity matrix $\bb{Q}$ \citep{albert1962estimating}, which is the limit of the transition probability matrix $\bb{P}$ as the time interval approaches zero. 
Suppose that $\delta_{it}$ is the number of pre-specified time units between $t^{th}$ and $(t-1)^{th}$ observation, then the transition intensity for subject $i$ from state $j$ to state $k$ at time $t$ is
\begin{equation}
    q_{jk}=\lim_{\delta_{it}\to 0} \frac{P(w_{it}=k|w_{i,t-\delta_{it}}=j)}{\delta_{it}}>0,j \neq k,
\end{equation}
and $q_{jj}=-\sum_{k\neq j}q_{jk}$. The corresponding transition probability matrix $\bb{P}(\delta_{it})$ can be calculated as the matrix exponential of $\delta_{it}*\bb{Q}$. The time intervals are assumed to be independent.

To investigate the impact of covariates on transition rates, the transition intensity matrix can be modeled through a log-linear model \citep{cook2002generalized,habtemichael2018missclassification}, such that
$\log(q_{jk}|\bb{x}_{it})=\bb{x}_{it}^{'}\bb{B}_{jk}$.
Although the CT-EHMFM is much more general than the DT-EHMFM, calculating the exponential of a matrix can be challenging. For simplicity, we approximate the exp($\bb{Q}$) using the $(\bb{I}+\bb{Q}/a)^{a}$ for some sufficiently large $a$ \citep{ross1996stochastic}. 


\section{Stabilized Expectation-Maximization Algorithm (SEMA)} \label{SEMA}
Let $\bb{\lambda} = (\{\bb{\mu}_{j}\}_{j=1}^{J}, \{\bb{\Lambda}_{j}\}_{j=1}^{J}, \bb{\Psi}, \{\bb{B}_{kj}\}_{k,j=1}^{J}, \bb{\pi})$. Given the sequence of latent states $\bb{W}_{i}$ and the latent factor scores $\bb{Z}_{i}$ for each $i$, and using Markov property of state sequence and the independence of $\bb{y}_{it}$ conditional on $w_{it}$, a joint probability distribution of the observations and all latent variables can be constructed as follows:
\begin{equation}
    L_{ci}(\bb{\lambda})
    =P(w_{i1})\times \prod_{t=2}^{T_{i}} P(w_{it}|w_{i,t-1},\bb{x}_{it}) \times  \prod_{t=1}^{T_{i}} P(\bb{y}_{it}|w_{it},\bb{z}_{it})P(\bb{z}_{it}),\label{likeli}
\end{equation}
which is also known as the complete likelihood function with full information for individual $i$. By the independence property of $\bb{Y}_{i}$, $\bb{W}_{i}$, and $\bb{Z}_{i}$ across $i$, the complete likelihood function ($L_{c}$) for the whole sample can be obtained by taking the product of equation(\ref{likeli}) over $i$. 

Our goal is to estimate $\bb{\lambda}$ by maximizing the likelihood function $L_{c}$, or its logarithm $l_{c}$. Since both $\bb{W}_{i}$ and $\bb{Z}_{i}$ are unobserved, the expectation-maximization (EM) algorithm is commonly used to identify the maximum likelihood estimator (MLE). As the name suggests, the EM algorithm finds a local maximum of the marginal likelihood by iteratively applying the expectation and maximization steps discussed below.

\subsection{Expectation Step (E-step)}  \label{E-step}
The E-step gets the expectation of $l_{c}$ given observations, with respect to the current conditional distribution of unobserved variables and the current parameter estimates $\bb{\lambda}^{v}$. Denote the target expectation (i.e., $E_{\bb{\lambda}^{v}}[l_{c}(\bb{\lambda})|\bb{Y},\bb{X}])$ as $\Omega(\bb{\lambda},\bb{\lambda}^{v})$. While an explicit form of the probability density function of $z_{it}$ exists, 
the calculation of conditional state probabilities can be computationally heavy. Therefore, we utilize a scaled version of the forward-backward algorithm (FBA) \citep{rabiner1989tutorial} to get the conditional state probabilities efficiently.

Specifically, we first define the forward probability $\alpha_{ij}(t)$ as $P(w_{it}=j|\bb{y}_{i1},\cdots,\bb{y}_{it})$. Denote $P_{j}(\bb{y}_{it})$ the probability density function of $\bb{y}_{it}$ given $w_{it}=j$ and $c_{i}(t)$ the conditional probability of observation $\bb{y}_{it}$ given all past observations. 
For each individual $i$ and state $j$, using a recursion scheme, the forward probabilities at $t=1,\cdots,T_{i}$ will be calculated as:
\begin{equation}
    \alpha_{ij}(1)=\frac{\pi_{j}P_{j}(\bb{y}_{i1})}{\sum_{j=1}^{J}\pi_{j}P_{j}(\bb{y}_{i1})}=\frac{\pi_{j}P_{j}(\bb{y}_{i1})}{c_{i}(1)};
\end{equation}
\begin{equation}
        \alpha_{ij}(t)=\frac{P_{j}(\bb{y}_{it})[\sum_{k=1}^{J}\alpha_{ik}(t-1)\bb{P}_{itkj}]}{\sum_{j=1}^{J}P_{j}(\bb{y}_{it})[\sum_{k=1}^{J}\alpha_{ik}(t-1)\bb{P}_{itkj}]}=\frac{P_{j}(\bb{y}_{it})[\sum_{k=1}^{J}\alpha_{ik}(t-1)\bb{P}_{itkj}]}{c_{i}(t)}.
\end{equation}
Then, we define the backward probability $\beta_{ij}(t)$ as $\frac{P(\bb{y}_{i,t+1},\cdots,\bb{y}_{i,T_{i}}|w_{it}=j,\bb{\lambda})}{c_{i}(t+1)}$. Similarly, we define a recursion form to update the backward probabilities at $t=T_{i},\cdots,1$ as follows:
\begin{eqnarray}
    \beta_{ij}(T_{i})=1,\textit{    } \beta_{ij}(t)=\frac{\sum_{k=1}^{J}\bb{P}_{i,t+1,jk}P_{k}(\bb{y}_{i,t+1})\beta_{ik}(t+1)}{c_{i}(t+1)}.
\end{eqnarray}
After that, in the smoothing step, denote $\epsilon^{v}_{ikj}(t)$ as $P(w_{i,t}=j,w_{i,t-1}=k|\bb{Y}_{i},\bb{\lambda}^{v})$ and $\gamma^{v}_{ij}(t)$ as $P(w_{it}=j|\bb{Y}_{i},\bb{\lambda}^{v})$. The target conditional state probabilities are functions of the forward probability and backward probability as follows:
\begin{eqnarray}
    \gamma^{v}_{ij}(t)=\alpha_{ij}(t)\beta_{ij}(t),\textit{    }
    \epsilon^{v}_{ikj}(t)=\frac{\alpha_{ik}^{v}(t-1)\bb{P}_{itkj}P_{j}(\bb{y}_{it})\beta_{ij}^{v}(t)}{c_{i}(t)}.
\end{eqnarray}
With the probabilities defined above, the $\Omega(\bb{\lambda},\bb{\lambda}^{v})$ can be written as the sum of three parts:
\begin{equation}
   \Omega(\bb{\lambda},\bb{\lambda}^{v})= constant + h(\bb{\pi})+h(\{\bb{B}_{kj}\}_{k,j=1}^{J}) -\frac{1}{2}h(\bb{\Psi},\{\bb{\Lambda}_{j}\}_{j=1}^{J},\{\bb{\mu}_{j}\}_{j=1}^{J}),
\end{equation}
where $h(\bb{\pi})$ depends on the initial state distribution, $h(\{\bb{B}_{kj}\}_{k,j=1}^{J})$ depends on the probability transition matrix, and $h(\bb{\Psi},\{\bb{\Lambda}_{j}\}_{j=1}^{J},\{\bb{\mu}_{j}\}_{j=1}^{J})$ is a function of parameters $\bb{\Psi}$, $\bb{\Lambda}_{j}$ and $\bb{\mu}_{j}$. Explicit forms are provided in Appendix B. Note that the E-step is identical for DT-EHMFM and CT-EHMFM, except for the dependence of $\bb{P}_{it,kj}$ on $\delta_{it}$ in CT-EHMFM.

\subsection{Maximization (M-step)}\label{M-step}
Within each M-step, since $h(\bb{\Psi},\{\bb{\Lambda}_{j}\}_{j=1}^{J},\{\bb{\mu}_{j}\}_{j=1}^{J})$, $h(\bb{\pi})$, and $ h(\{\bb{B}_{kj}\}_{k,j=1}^{J})$ do not share parameters, we maximize each of them separately. 
The estimator of $\bb{\pi}$, $\bb{\Lambda}_{j}$, $\bb{\mu}_{j}$, and $\bb{\Psi}$ can be directly derived by setting $h(\bb{\pi})=0$ and $h(\bb{\Psi},\{\bb{\Lambda}_{j}\}_{j=1}^{J},\{\bb{\mu}_{j}\}_{j=1}^{J})=0$ (see Appendix B for details). 
For $\{\bb{B}_{kj}\}_{k,j=1}^{J}$, 
a one-step Newton-Raphson (NR) algorithm is implemented. 

First, considering the DT-EHMFM, let $\bb{S}_{kj}$ be the first-order partial derivative of $ h(\{\bb{B}_{kj}\}_{k,j=1}^{J})$ with respect to $\bb{B}_{kj}$ and the $(j,j^{'})$ block entry of $\bb{M}_{k}$ ($\bb{M}_{k}(j,j')$) to be the negative second-order partial derivative with respect to $\bb{B}_{kj}$ and $\bb{B}_{kj^{'}}$. 
 Let $\bb{S}_{k}$ and $\bb{B}_{k}$ be defined similarly as  $\bb{S}_{k}=(\bb{S}_{k1}^{'},\cdots,\bb{S}_{k,k-1}^{'},\bb{S}_{k,k+1}^{'},\cdots,\bb{S}_{kJ}^{'})^{'}$. Then $\bb{B}_{k}$ is updated as $\bb{B}_{k}^{v+1}=\bb{B}_{k}^{v}+\bb{M}_{k}^{-1}\bb{S}_{k}$. 
Alternatively, to ensure the stability of the algorithm and control the distance between $\bb{B}_{k}^{v+1}$ and $\bb{B}_{k}^{v}$, we may update $\bb{B}_{k}$ as $\bb{B}_{k}^{v+1}=\bb{B}_{k}^{v}+(\bb{M}_{k}+\bb{S}_{k}^{T}\bb{S}_{k})^{-1}\bb{S}_{k}$.

Second, considering the CT-EHMFM, though the corresponding likelihood function is in the same form as that corresponding to the DT-EHMFM, the maximization step for CT-EHMFM is more complicated as it involves operations of the matrix exponential. Let $\bb{\theta}$ be an ordered vector of all the transition model parameters, 
such that $\bb{\theta}=vec(\{\bb{B}_{kj}^{'}\},k\neq j)$. Recalling the first derivative of the matrix exponential \citep{zhou2020continuous} 
and using Theorem 1 in \citet{van1978computing}, 
\begin{eqnarray}
 \frac{\partial}{\partial \theta_{u}} exp(\bb{A}(\theta_{u}))  =  \int_{0}^{1} exp(u\bb{A}(\theta_{u})) \tilde{\bb{A}}(\theta_{u}) exp((1-u)\bb{A}(\theta_{u}))du=  exp(\bb{H})_{0:J,J:2J},\nonumber
\end{eqnarray}
where $\tilde{\bb{A}}(\theta_{u})=(\tilde{A}_{ij}(\theta_{u}))=(\frac{\partial A_{ij}(\theta_{u})}{\partial \theta_{u}})$ and $\bb{H} = \begin{bmatrix}
\bb{A}(\theta_{u}) & \tilde{\bb{A}}(\theta_{u})\\
\bb{0} & \bb{A}(\theta_{u}) 
\end{bmatrix}$. Denote $\frac{\partial \bb{P}_{kj}(\delta_{it})}{\partial \theta_{u}}$ as the $(k,j)$ entry of the first derivative of $\bb{P}(\delta_{it})$ with respect to $\theta_{u}$ (i.e., the $u^{th}$ entry of $\bb{u}$). Having the first derivative of $exp(\delta_{it}*\bb{Q}) = \bb{P}(\delta_{it})$ with respect to each component of $\bb{\theta}$ calculated accordingly, a variant of NR, the Fisher scoring algorithm (FS) \citep{kalbfleisch1985analysis} can be directly implemented to update the parameter vector $\bb{\theta}$ to forbid the calculation of the second derivative of matrix exponential. Specifically, denote $\bb{S}^{*}$ be the score function, and $\bb{S}^{*}_{u}$ be the $u^{th}$ entry of the score function $\bb{S}^{*}$. Then,
\begin{equation}
    \bb{S}^{*}_{u}(\bb{\theta})=\frac{\partial h(\{\bb{B}_{kj}\}_{k,j=1}^{J})}{\partial \theta_{u}}= \sum_{i=1}^{N}\sum_{t=2}^{T_{i}}\sum_{j=1}^{J}\sum_{k=1}^{J}\frac{\epsilon^{v}_{ikj}(t)}{\bb{P}_{kj}(\delta_{it})}\frac{\partial \bb{P}_{kj}(\delta_{it})}{\partial \theta_{u}}.
\end{equation}
Let $\bb{M}^{*}$ be the negative Fisher information matrix. Its $(u,v)$ entry $\bb{M}^{*}_{uv}$ is in the form of:
\begin{equation}
    \bb{M}^{*}_{uv}(\bb{\theta})=\sum_{i=1}^{N}\sum_{t=2}^{T_{i}}\sum_{j=1}^{J}\sum_{k=1}^{J}\frac{\gamma^{v}_{ik}(t-1)}{\bb{P}_{kj}(\delta_{it})}\frac{\partial \bb{P}_{kj}(\delta_{it})}{\partial \theta_{u}}\frac{\partial \bb{P}_{kj}(\delta_{it})}{\partial \theta_{v}}.
\end{equation}
After getting both the score function and the Fisher information matrix, parameters $\bb{\theta}$ can be updated as $\bb{\theta}^{v+1}=\bb{\theta}^{v}+\bb{M}^{*}(\bb{\theta}^{v})^{-1}\bb{S}^{*}(\bb{\theta}^{v})$. 
Similar to the DT-EHMFM, a stabilized version is employed in practice 
with $\bb{\theta}^{v+1}=\bb{\theta}^{v}+\{\bb{M}^{*}(\bb{\theta}^{v})+\bb{S}^{*}(\bb{\theta}^{v})^{T}\bb{S}^{*}(\bb{\theta}^{v})\}^{-1}\bb{S}^{*}(\bb{\theta}^{v}).$ 

The complete iterative algorithm is summarized in Appendix C. Note that the algorithm requires the specification of (K, J), which are typically unknown in practice. In this study, we propose to determine (K, J) using information criteria, the efficacy of which is evaluated in Section \ref{simulation3}.  




\section{Simulation Study}\label{Simulation}
This section conducts a simulation study to evaluate the proposed methods using synthetic data designed to resemble the AURORA data. The simulations are under similar settings as the AURORA data with respect to the sample size (\emph{N}), number of observations per subject ($T_{i}$), number of response variables (\emph{p}), and the number of covariates (\emph{d}) in the transition model. Specifically, the synthetic data is generated randomly with \emph{N}=200, \emph{p}=23, and \emph{d}=3. $T_i = 10$ for discrete-time (DT) setting, while $T_i \in[50,100]$ for the continuous-time (CT) setting. Furthermore, we assume that $J=3$ and $K=3$. See Appendix D for the complete data generation process and the true values of parameters. In the following, Subsection \ref{simulation1} evaluates the reliability of the proposed model by comparing the empirical results of parameter estimates with their respective true values. Comparing the performance of the proposed method to that of baseline methods, Subsection \ref{simulation2} demonstrates the benefit of integrating the EFM and the feature-based TM with the standard HMMs. Finally, Subsection \ref{simulation3} explores the performance of information criteria in model selection.

\subsection{Simulation 1}\label{simulation1}


To validate the estimation procedure, we implement the SEMA under the assumption that both $J$ and $K$ are known a priori. Initial values for parameters are determined by first fitting Gaussian Mixture Models (GMM) and then fitting EFM for each estimated group. Guided by the insights from a pilot study, we set the maximum number of iterations for each replication at 100. The reliability and precision of the proposed methods are then evaluated from two perspectives: i) the accuracy of each individual parameter estimate and ii) the misclassification rate ($C_{mis}$), which quantifies the proportion of estimated states that diverge from the actual states.

The accuracy of $\bb{\pi}$, $\bb{\mu}$, $\bb{\Lambda}$, and $\bb{\Psi}$ is determined by calculating the average absolute difference (AAD) between parameter estimates and their true values. Mathematically, the AAD of a parameter matrix $\bb{o}$ is expressed as $AAD(\bb{o})=\frac{\sum_{i=1}^{r}|\hat{o_{i}}-o_{i}|}{r}$,
where $o_{i}$ is a single entry in the matrix $\bb{o}$ and $r$ denotes the total number of free parameters in the parameter matrix $\bb{o}$. The mean of AADs (standard errors in the parentheses) aggregated over $100$ random seeds are presented in Table \ref{tab:PPE}. For both the CT-EHMFM and the DT-EHMFM, the mean ADDs of all parameter matrices are sufficiently close to zero with small standard errors, suggesting a good parameter recovery. 
\begin{table}
\caption{The Mean (standard error) AADs of $\bb{\pi}$, $\bb{\mu}$, $\bb{\Lambda}$, and $\bb{\Psi}$, and $C_{mis}$ of the estimations.  \label{tab:PPE}}
\begin{center}
\begin{tabular}{rrrrrr}
Parameter & $\bb{\pi}$ & $\bb{\mu}$ & $\bb{\Lambda}$ & $\bb{\Psi}$ & $C_{mis}$\\\hline
DT-EHMFM & .027(.014) & .040(.005) & .037(.002) & .030(.005) & .0023(.0010) \\
CT-EHMFM & .026(.013) & .015(.002) & .014(.001) & .011(.002) & .0024(.0005)\\\hline
\end{tabular}
\end{center}
\end{table}
In Table 2, we present the mean bias (standard error in parenthesis) of each parameter in the transition model. The mean bias of each parameter in the transition model is close to zero for both CT-EHMFM and DT-EHMFM. Nonetheless, the standard errors for each parameter estimate in the transition model of CT-EHMFM are considerably smaller than those of the DT-EHMFM, which is primarily attributable to the longer panel lengths $T_i$. In the DT-EHMFM setting, each subject has only ten observations, whereas each subject has at least 50 observations in the CT-EHMFM setting. Additional simulations revealed that $T_{i}$ is a critical factor influencing the parameter estimation, which will be illustrated later.

Moreover, we present the mean (standard deviation) of $C_{mis}$ in the last column of Table \ref{tab:PPE}. On average, only 0.24\% (0.0005) and 0.22\% (0.0010) of observations are misclassified under the CT-EHMFM and DT-EHMFM settings, respectively, demonstrating the outstanding performance of the SEMA algorithm in estimating latent states.

Intuitively, various factors, including sample size (N), the number of measurements per individual ($T_i$), the sizes of $J$ and $K$, the size of the common variance $\bb{\Psi}$, the differences in $\bb{\mu}_{j}$ and $\bb{\Lambda}_{j}$ between states, and the frequency of state transitions, can affect the performance of parameter estimation. Additional simulations for both DT-EHMFM and CT-EHMFM in Appendix E.2 reveals that estimation performance for $\bb{\mu}$, $\bb{\Lambda}$, $\bb{\Psi}$, and $\bb{B}$, and the rate of correct classification are improved when (i) common variances decrease, (ii) differences in $\bb{\mu}_{j}$ and $\bb{\Lambda}_{j}$ between states increase, (iii) $J$ decreases, or (iv) sample size ($N$) or panel length ($T_{i}$) increases. Increasing the size of $K$ or using a $\bb{B}$ that induces infrequent transitions has little effect on the estimation of the majority of parameters, but it enhances the precision of transition probability estimation, thereby reducing the misclassification rate. The estimation of $\bb{\pi}$ is improved solely by increasing sample size ($N$) or state-to-state differences in $\bb{\mu}_{j}$. 

\subsection{Simulation 2}\label{simulation2}
This section compares the performance of the proposed methods and the baseline approaches in correctly identifying latent states.  Three benchmark methods are under our consideration: i) TM+independent HMM, which assumes independence among observed features given the states; ii) CFM+TM+HMM, which addresses interrelationships but inaccurately pre-specifies the latent factor structure by setting certain loading matrix entries to zero; and iii) EFM+HMM, which assumes a homogeneous transition probabilities matrix for all subjects. We first repeat the data generation process of Simulation 1. Then, we consider three additional scenarios by adjusting the state-to-state differences in $\bb{\mu}_j$ to be closer ($\bb{\mu}$: medium diff), increasing the similarity of the $\bb{\Lambda}_j$ at different states ($\bb{\Lambda}$: medium diff), and increasing the significance of the covariance matrix $\bb{\Psi}$ ($\bb{\Psi} = 2 \times I$), respectively.

\begin{figure}
    \centering
    \includegraphics[width=5.5in]{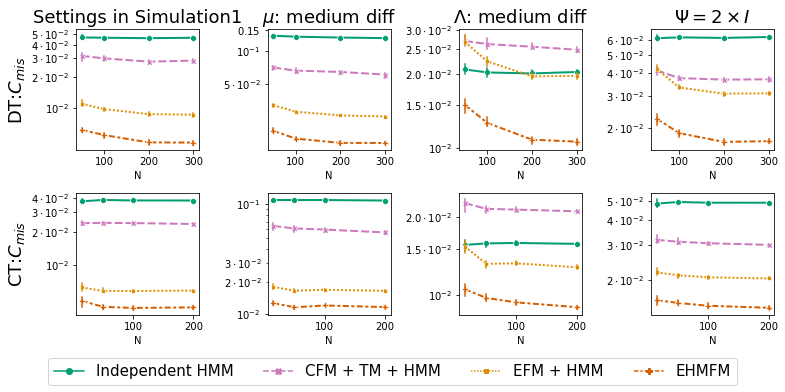}
    \caption{$C_{mis}$ of various methods. The error bars represent the 95\% CI. For the DT setting, $T=10$. For the CT setting, $50\leq T\leq100$. The first column shows the results under the settings we used in simulation 1. The last three columns summarize the results under different settings by varying the true value of $\bb{\mu}$, $\bb{\Lambda}$, and $\bb{\Psi}$, respectively. The true values of $\bb{\mu}$ and $\bb{\Lambda}$ with a medium state-to-state difference can be found in Appendix E.2.}
    \label{simu2_comparison_N}
\end{figure}

As depicted in Figure \ref{simu2_comparison_N}, our proposed methods (CT-EHMFM and DT-EHMFM) consistently outperform the benchmark methodologies in both CT and DT settings. Regardless of sample size, our methods consistently achieve the lowest misclassification rate, nearly approximating zero, thereby emphasizing the importance of each component in our proposed models. Specifically, the comparison with TM+independent HMM shows the importance of accounting for the interrelationship between observed features; the comparison with CFM+TM+HMM reveals the risk of incorrectly specifying the interrelationship structure; and the comparison with EFM+HMM demonstrates the inadequacy of assuming homogeneous transition probabilities.


\subsection{Simulation 3}\label{simulation3}
Information criteria such as the Akaike information criteria (AIC) and the Bayesian information criteria (BIC) have been widely used in model selection \citep{preacher2013choosing,Song2017}. Within this simulation study, we investigate whether the AIC or BIC is reliable for determining J and K simultaneously. We repeat the data generation process of Simulation 1, but implement the proposed methods with a different set of $(J, K)$ for each replicate when fitting the generated data. Let $J=\{2,3,4\}$ and $K=\{2,3,4\}$. We consider all possible combinations of $J$ and $K$, yielding a total of nine fitted candidate models for each replicate. 


Table \ref{tab:BIC/AIC} presents the results of $100$ replications, suggesting that both BIC and AIC performed well in model selection. In the simulation study for the DT-EHMFM, AIC recommends a model with the correct $J$ and $K$ in 94\% of replications, while BIC yields the accurate recommendation in 100\% of replications. Notably, as the total number of observations increases, AIC's performance will improve  (see related results in Appendix E.3). In the case of the CT-EHMFM, both AIC and BIC consistently recommend the model with accurate $J$ and $K$. Therefore,  we believe that the sample size and the number of observations per individual in the processed AURORA data will yield reliable information criteria-based model selection results and, consequently, reliable parameter estimation.

\begin{table}
\caption{The percentage of (J, K) pairs selected based on AIC/BIC.  \label{tab:BIC/AIC}}
\begin{center}
\begin{tabular}{rrrrr}
 & J & K & Percentage (DTE) & Percentage (CTE) \\\hline
AIC & 3 & 3 & 94\% & 100\% \\
 & 4 & 3 & 6\% & - \\
BIC & 3 & 3 & 100\% & 100\% \\\hline
\end{tabular}
\end{center}
\end{table}
\section{Analysis of the AURORA Data} \label{real_ana}


Due to the inherent irregularity in the collection of mobile device data, we apply the more general method CT-EHMFM to the smartwatch data from the AURORA study. We consider a collection of 54 candidate models ($J=\{1,2,\cdots,6\}$; $K = \{1,2,\cdots,9\}$). For each candidate model, the SEMA algorithm is implemented with multiple random seeds, and the seed yielding the highest estimated likelihood is selected. Then, we use AIC and BIC to compare all fitted candidate models with different J and K. Finally, a model with three states $(J=3)$ and eight factors per state $(K=8)$ is suggested. In the following subsections, we focus on the interpretation of parameter estimates and biological findings from four perspectives: i) the interpretation of three estimated states, ii) the co-occurring patterns of symptoms, iii) the relationship between transition probability and demographic factors such as age and gender, and iv) the common structure of the loading matrix.

\subsection{Interpretation of Hidden States}
\begin{figure}
\begin{center}
\includegraphics[width=6in]{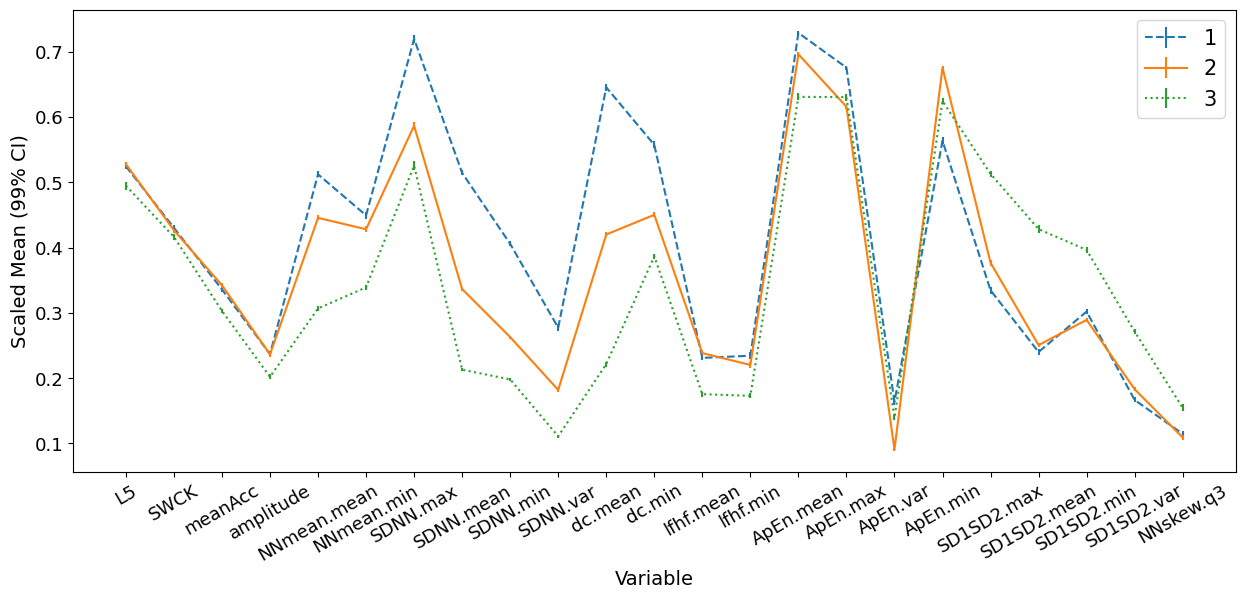}
\end{center}
\caption{Relative sample mean for features in each estimated states. The error bars represent the 99\% CI, which are small and hence hard to distinguish.\label{fig:Feature}}
\end{figure}

To investigate the biological differences between different states, we first focus on the selected features. Figure \ref{fig:Feature} depicts the scaled 
sample means of each feature across different states, along with the corresponding $99\%$ confidence interval (CI). Further pairwise Tukey tests indicate significant differences between states concerning almost all features, with the exception of state $1$ and state $2$ when concerning amplitude, SWCK, L5, and NNskew.q3.

Overall, features related to average heart rate (NNmean-related features), heart rate variability (SDNN-related features), and heart deceleration capacity (dc-related features) vary significantly between the three latent states. The values of these features show a sequential decrease from state $1$ to state $2$, and then to state $3$. According to previous research, lower heart rate variability and deceleration capacity are associated with a higher mortality rate \citep{kleiger2005heart,kantelhardt2007phase}. Therefore, it is reasonable to conclude that latent states 1 to 3 represent participants' health in descending order, with state 1 being the healthiest and state 3 being the least healthy. Moreover, regarding activity features, states $1$ and $2$ have similar but higher means compared to state $3$, indicating that participants in states $1$ and $2$ have higher levels of daily activity and thus are in better health than those in state $3$. 

Among all the features related to heart rate randomness or unpredictability (lfhf, ApEn, and SD1SD2), state $3$ demonstrates significantly higher values for SD1SD2-related features but lower values for lfhf-related features compared to state $1$ and $2$, suggesting a different interpretation of the estimated states than our previous interpretation. However, it is important to note that previous studies have suggested that the relationship between these features and the psychological or physiological state is neither straightforward nor unique \citep{von2017resolving}.

\begin{figure}
\begin{center}
\includegraphics[width=6in]{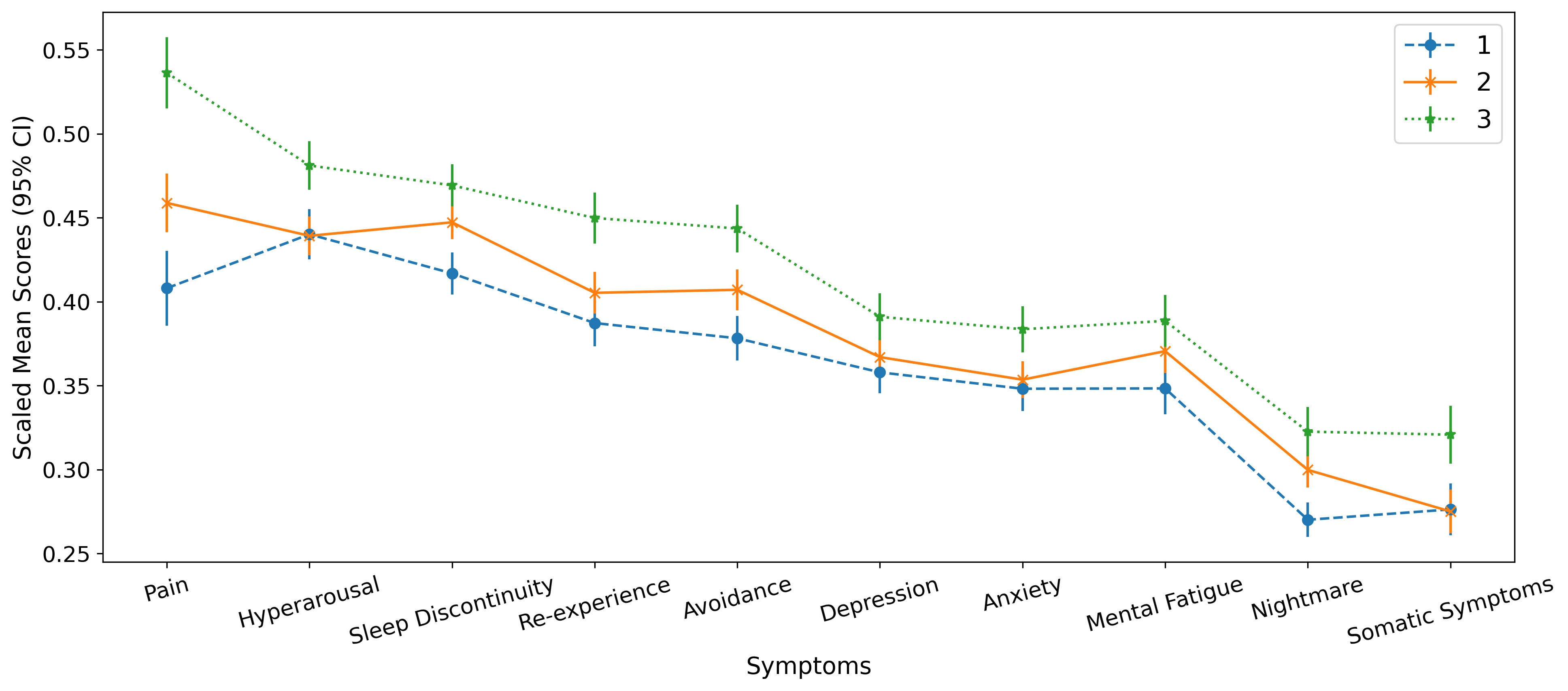}
\end{center}
\caption{Sample mean for each symptom in each estimated state. The error bars represent the 95\% CI.\label{fig:Survey}}
\end{figure}

To confirm the validity of the three states, we further compare their differences regarding self-report symptoms collected from a flash survey (details are provided in Appendix G). Based on the RDoC framework, ten latent constructs associated with APNS were developed using flash survey items selected by domain experts: Pain, Loss, Sleep Discontinuity, Nightmare, Somatic Symptoms, Mental Fatigue, Avoidance, Re-experience, and Anxious. Retaining only observations for each individual whose estimated states are known on the same day they submitted survey responses, we summarized the flash survey data with means and $95\%$ CIs in Figure \ref{fig:Survey}. While $0$ represents the least severity, $1$ represents the greatest severity. 

Overall, state $1$ exhibits the lowest severity level for all ten symptoms, while state $3$ has the highest severity level. Based on the Tukey tests, while states $1$ and $2$ are not statistically different in hyperarousal, re-experience, anxiety, and somatic symptoms, they diverge significantly from state $3$ in these constructs. While the differences in nightmare and sleep discontinuity between states 3 and 2 are not significant, they are statistically more severe than in state 1. For mental fatigue and depression, only the difference between state $1$ and state $3$ is statistically significant. 

In summary, both the flash survey data and the AURORA data (HRV, Activity) support our interpretation of the three latent states. State 1 is the healthiest, while state 3 indicates having the most severe APNS symptoms.

\subsection{Co-occurring Pattern of Symptoms}
When studying the pattern of co-occurring symptoms within each hidden state, we limit our attention to observations collected during the first week. For each estimated state,  the correlations between all ten symptoms are calculated. In state $1$ (relative health state), there is a high degree of correlation between hyperarousal and anxiety, which implies that if a patient in state $1$ experiences severe hyperarousal symptoms, it is highly likely that he or she will also suffer from severe anxiety symptoms. In other words, there is a high likelihood of concurrent manifestation of hyperarousal and anxiety in patients in state 1. In state 2, symptoms typically do not co-occur due to the lack of a high correlation between any pair of symptoms. In state 3 (the state with more severe disorders), symptoms such as depression, hyperarousal, anxiety, and re-experience are more likely to co-occur.


\subsection{Transition Probability}
\begin{figure}
\begin{center}
\includegraphics[width=6in]{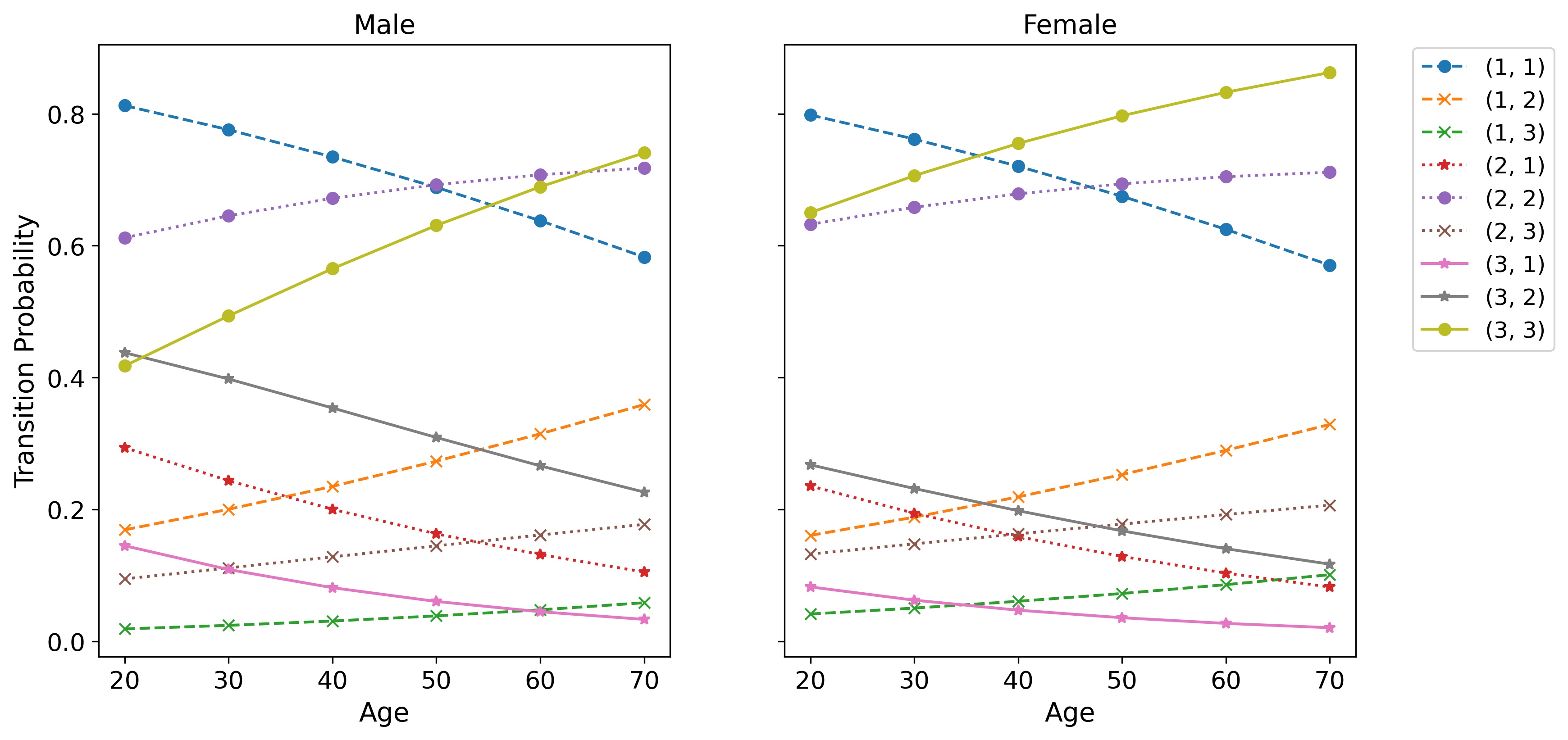}
\end{center}
\caption{Estimated transition probability. Fix $\delta_{it} = 1$. (a, b) indicates a transition from state a to state b.\label{fig:tran_prob}}
\end{figure}

This section mainly investigates the heterogeneity of 1-day transition probabilities among subjects by analyzing the transition probabilities with a time interval $\delta_{it}=1$. We estimated the transition probabilities for males and females within the sample age range, as depicted in Figure \ref{fig:tran_prob}. Lines embellished with circles illustrate the probability of remaining in the same state, lines adorned with stars indicate the likelihood of transitioning to a more severe state, while lines marked with `x' reflect the chance of improvement in psychological conditions. 

Overall, both males and females have a tendency to remain in their current state, with infrequent state transitions, aligning with most existing literature. For the male group, the probability of staying at states $3$ and $2$ increases with age, while the probability of staying at state $1$ decreases as aging. Moreover, while the likelihood of psychological deterioration increases with age, the chance of psychological state improvement decreases as age increases. Specifically, while the probability of transitioning from the most severe state (state 3) to the healthiest state (state 1) approaches zero as age increases, the likelihood of the reverse transition approaches zero as age decreases, with the direct transition between state $1$ and state $3$ being the least likely. The female group exhibits a similar trend to the male group, with the notable exception that females have a greater likelihood of remaining in the most severe state (state 3) compared to males.

In summary, our analysis of the AURORA data suggests that older patients are more likely to transition to a more severe psychological state. Moreover, achieving psychological improvement becomes increasingly challenging as one ages. 


\subsection{Factor Loading Structure}
Finally, we are interested in the structure of the factor loading matrix that explains the interrelationships of observed features within each state. To facilitate the interpretation, the estimated loading matrix presented in Appendix F is rotated by the promax rotation \citep{hendrickson1964promax,browne2001overview}, and then standardized by the estimated standard deviation of each variable ($\bb{\Lambda}_{j}\bb{\Lambda}_{j}^{'}+\bb{\Psi}$). Factor loadings with absolute values greater than $.4$ are considered to indicate moderate to high correlations between a feature and a factor \citep{peterson2000meta}, and are bolded.

Overall, factor loading matrices for the three states share some similarities but also have distinct differences, implying heterogeneous interrelationship structures between different states. For all states, factor $0$ is defined by features related to heart rate variability and irregularity. While the structure of factor $0$ in $\bb{\Lambda}_{2}$ and $\bb{\Lambda}_{3}$ is identical, which is defined by SDDN-, dc- and ApEN-related features, factor $0$ in $\bb{\Lambda}_{1}$ is solely defined by SDNN, and ApEn, with dc-related features contribute to the factor $7$ in $\bb{\Lambda}_{1}$. Similarly, the components of factor $2$ are consistent in states $2$ and $3$, consisting of features related to lfhf and SD1SD2. However, factor $2$ in state $1$ includes two additional SD1SD2-related statistics that define the factor $5$ in both states $2$ and $3$. 

Factor $1$ is defined by ApEn-related statistics for all states, but the weighting of each statistic varies across states. Factor $3$ is positively correlated with the mean heart rate (NNmean) and negatively correlated with the skewness of heart rate (NNskew). Factor $4$ summarizes activity features, which shows a negative correlation between average activities (i.e., meanAcc, amplitude, and L5) and the number of transitions between wake and sleep (SWCK), suggesting that individuals who engage in more daytime activities tend to have better sleep quality. Factor $6$ is defined solely by the summary statistics of SDNN.

In summary, while the factor loading matrix does not differ significantly between states, state distinction in means contributes the most to distinguishing between states in this case study. 

\section{Conclusions and Discussion} \label{conclusion}
This paper investigates the unique challenges of analyzing longitudinal mobile health data, including interdependent variables with unknown interrelationship structures, heterogeneous transition probabilities, and irregular measurements. To address these issues, we propose two HMM-based models, the DT-EHMFM and the CT-EHMFM, for multivariate longitudinal data collected regularly and irregularly, respectively. Furthermore, the performance of the corresponding Stabilized Expectation-Maximization algorithm for maximum likelihood estimation is supported by extensive simulation studies. Finally, we analyzed the AURORA data and drew biological findings comparable with previous research, implying that the mobile health data sourced from consumer-grade devices, together with the proposed methods, have the immense potential to facilitate mental health diagnostics and understand the dynamic transition mechanism.

The proposed methods can be extended in several ways. First, most entries in the estimated factor loading matrix are close to zero, indicating sparse factor loading matrices in real analysis. Although various methods (e.g., factor rotations and setting factor loadings below specific cutoffs to 0) are frequently used to simplify interpretation, the choice of these methods is subjective. The sparse exploratory factor loading analysis \citep{xie2010penalized, chen2012sparse} provides an automated approach to set the loading entries of redundant variables to $0$, thereby enhancing the interpretability of loading matrices without reliance on subjective factors. Therefore, incorporating sparse regularization into the factor loading matrix is an important extension of our current work worth studying. 

Second, a large number of baseline covariates are typically available in real data. However, we have no prior knowledge about the significance of each covariate in determining the transition probability. Hence, integrating regularization into the transition model to assist with variable selection can be extremely useful.

Third, mental health, according to domain knowledge, is exceptionally diverse. A key assumption of the HMM is the independence between $\bb{Y}_{it}$s given the hidden states. Intuitively, it is easy to be violated in real applications, especially given the likelihood of autocorrelation between observations collected from the same subject. Therefore, adding a random effect to the current model to account for the inter-patient heterogeneity is a natural extension \citep{altman2007mixed,Song2017}.


Finally, previous HRV-related studies are often conducted in well-controlled laboratory environments. Thus, all existing HRV feature extraction tools rely on resting-state heart rate data. However, heart rate data collected in open environments will inevitably contain additional noise. For example, it is reasonable to expect that HRV features corresponding to different activity states (e.g., exercising and resting) would differ significantly. Therefore, recognizing the lack of tools to extract HRV features corresponding to different activity states, we believe it would be advantageous to develop a preprocessing pipeline to concurrently process heart rate and activity data to derive appropriate HRV features.

\newpage
\setcounter{equation}{0}
\renewcommand{\theequation}{\thesection.\arabic{equation}}

\bibliographystyle{apalike.bst}
\bibliography{Reference_main}

\newpage

\appendix
\setcounter{equation}{0}
\renewcommand{\theequation}{A.\arabic{equation}}

\setcounter{table}{0}
\renewcommand{\thetable}{A.\arabic{table}}


\section*{Appendix A: Description of AURORA Data} \label{Appendix:0}

In the AURORA study \citep{mclean2020aurora}, trauma survivors aged 18-75 presenting to participating EDs within 72 hours of trauma exposure were screened for enrollment eligibility. Motor vehicle collisions (MVC), physical assault, sexual assault, falls $>$10 feet, or mass casualty incidents are automatically qualified for enrollment. Major exclusion criteria include administration of general anesthesia, long bone fractures, laceration with significant hemorrhage, visual or auditory impairment precluding completion of web-based neurocognitive evaluations and/or telephone follow-ups, prisoners, pregnant or breastfeeding, and ongoing domestic violence. Proficiency in written and spoken English and owning an internet-accessible iOS/Android smartphone were also prerequisites. 
Participants used in this study are from the third data freeze, which includes those who were enrolled at least up to day 67 of the study. Participants who became pregnant, were incarcerated, or died during the duration of the study are excluded. 

Prior research has suggested that heart rate variability (HRV) and activity features are associated with APNS \citep{hartmann2019heart,jung2019heart}. Therefore, our focus is primarily on HRV and activity features extracted from PPG and accelerometer data collected from Verily's smartwatches during the first 100 days post-enrollment. Activity features are extracted on a 24-hour window to evaluate the daily activity patterns of the participants. After converting accelerometer data to activity counts, the mean and standard deviation of activity counts for each 24-hour interval are calculated. Additionally, cosinor rhythmometry features were derived to capture circadian rhythm. HRV features were derived by first calculating the beat-to-beat (BB) interval \citep{shaffer2017overview} time series from PPG data. After identifying and removing noises from the BB interval time series, normal-to-normal (NN) interval time series are derived. Finally, the NN interval time series was analyzed using a $5$-minute window with a $30$-second sliding step to derive HRV features. In the following subsections, we will discuss the activity data and HRV data in more detail before concluding with a summary of the final dataset of interest. 

\subsection*{Appendix A.1: Activity Features}\label{activity}
There are four activity features considered in this study. The {meanAcc} is the mean of the activity counts calculated by the approach described in \citet{borazio2014towards}, serving as descriptive statistics about the level of activity in the given time period. The {amplitude} is a feature derived from the Cosinor Rhythmometry Analysis \citep{cornelissen2014cosinor} to quantify circadian rhythm. By applying the Cole-Kripke algorithm \citep{cole1992automatic} on accelerometry epochs, each epoch is classified as either wake or sleep. The {SWCK} is the number of transitions between wake and sleep epochs divided by the length of the data. Based on the raw accelerometer data \citep{van1999bright}, the average activity over the five least active hours ({L5}) is calculated, indicating nighttime activity.

\subsection*{Appendix A.2: Heart Rate Features} \label{HRV}

Technically, HRV features can be grouped into three categories: time-domain measures, frequency-domain measures, and nonlinear measures \citep{shaffer2017overview}. For this study, seven heart rate characteristics were chosen to assess the mean, variability, and complexity of the heart rate time series. {NNmean} is the average heart rate, while the {NNskew} and {SDNN} are the skewness and standard deviation of the NN interval \citep{shaffer2017overview}, respectively. In particular, higher skewness indicates rapid accelerations or decelerations. The ratio of low-frequency power to high-frequency power is denoted by {Lfhf}. A low {Lfhf} ratio suggests parasympathetic dominance (i.e., engage in tend-and-befriend behaviors), whereas a high {Lfhf} ratio shows sympathetic dominance (i.e., engage in fight-or-flight behaviors) \citep{shaffer2017overview}. According to \citet{kantelhardt2007phase}, {DC} is a predictor of mortality in heart attack survivors. The lower the {DC} index, the greater the mortality risk. The remaining two variables characterize the time between successive heartbeats (R-R interval). While {SD1SD2} assesses the unpredictability of an R-R interval time series, {ApEn} measures its regularity and complexity. A large {ApEn} shows R-R interval volatility, whereas a small ApEn indicates a steady and predictable temporal sequence of R-R intervals \citep{shaffer2017overview}. 
To align with the activity data, daily statistical summaries of each HRV feature, such as mean, median, minimum, maximum, kurtosis, skewness, interquartile range, and standard deviation, are used.

\subsection*{Appendix A.3: Data Pre-Processing}
We consider a subset of the AURORA data by selecting patients involved in MVC before presenting to the ED to investigate the dynamic change in patients' mental health conditions in the $100$ days following MVC exposure. We maintain only observations with no missing activity data and a positive wake percentage. Regarding heart rate data, an individual ideally has 2880 records per day. We keep only observations for days with at least 30\% (equivalent to 2880*0.3) recordings in order to derive representative daily summary statistics. Before fitting our model, we further apply the Box-Cox transformation \citep{osborne2010improving} to each variable to reduce the skewness of the original data, eliminate outliers, and standardize the data by dividing each variable by its sample standard deviation.



The final dataset consists of observations from 258 patients, with each patient's total number of records ranging from $17$ to $99$. In total, there are $23$ variables of interest, of which four are derived from the activity data, and $19$ are derived from the HRV data. Besides, the transition model considers age (range from $18$-$74$) and gender ($0$-male, $1$-female) to examine their influence on transition probabilities. Note that observations in this dataset are irregularly sampled and hence suitable for analysis using the CT-EHMFM. Alternatively, researchers can employ the proposed DT-EHMFM model by selecting a subset of equally-spaced observations. The largest aligned subset for the AURORA dataset consists of $180$ patients, each with observations collected on \{day $2$, day $12$, $\cdots$, day $82$\}. 

\section*{Appendix B: Technical Details} \label{Appendix:1}
In this section, we show the explicit forms of the components in the $\Omega(\lambda,\lambda^{v})$ and the explicit forms of the MLE of $\bb{\pi}$, $\Lambda$, and $\bb{\Psi}$.

\subsection*{Appendix B.1: Supplement for E-step} \label{E-add}
Denote $\widetilde{\bb{\Lambda}_{j}}=(\bb{\Lambda}_{j} , \bb{\mu}_{j})\in \mathcal{R}^{p\times(K+1)}$ and $\widetilde{\bb{z}_{it}}=(\bb{z}_{it}^{T},1)^{T} \in \mathcal{R}^{(K+1)}$. Each of the three parts has an explicit form as follows:
\begin{equation}
    h(\bb{\pi})=\sum_{i=1}^{N}\sum_{j=1}^{J}\gamma^{v}_{ij}(1)log(\pi_{j}),\label{A}
\end{equation}

\begin{equation}
    h(\{\bb{B}_{kj}\}_{k,j=1}^{J})=\sum_{i=1}^{N}\sum_{t=2}^{T_{i}}\sum_{j,k=1}^{J}\epsilon^{v}_{ikj}(t)log(\bb{P}_{itkj}),\label{B}
\end{equation}
\begin{equation}\label{C}
\begin{split}
     h(\bb{\Psi},\{\bb{\Lambda}_{j}\}_{j=1}^{J},\{\bb{\mu}_{j}\}_{j=1}^{J})=\sum_{i=1}^{N}\sum_{t=1}^{T_{i}}\sum_{j=1}^{J} &\gamma^{v}_{ij}(t)log|\bb{\Psi}|\\&+\gamma^{v}_{ij}(t)\bb{y}_{it}^{'}\bb{\Psi}^{-1}\bb{y}_{it}-2\gamma^{v}_{ij}(t)\bb{y}_{it}^{'}\bb{\Psi}^{-1}\widetilde{\bb{\Lambda}_{j}}
        E_{\bb{\lambda}^{v}}(\widetilde{\bb{z}_{it}}|\bb{y}_{it},w_{it})\\&+\gamma^{v}_{ij}(t)tr(\widetilde{\bb{\Lambda}_{j}}^{'}\bb{\Psi}^{-1}\widetilde{\bb{\Lambda}_{j}}E_{\bb{\lambda}^{v}}(\widetilde{\bb{z}_{it}}\widetilde{\bb{z}_{it}}^{'}|\bb{y}_{it},w_{it})).
\end{split}
\end{equation}
Note that, in the discrete-time case, the $log(\bb{P}_{itkj})$ in the equation (\ref{B}) can be further expressed as $\bb{x}_{it}^{T}\bb{B}_{kj}-log({\sum_{l=1}^{J}e^{\bb{x}_{it}^{T}\bb{B}_{kl}}})$.
Using the Woodbury matrix identity \citep{rasmussen2003gaussian}, let $\bb{M}_{j}^{v}= (I+\bb{\Lambda}_{j}^{v'}\bb{\Psi}^{v^{-1}}\bb{\Lambda}_{j}^{v})^{-1}$, two expectation terms in $h(\bb{\Psi},\{\bb{\Lambda}_{j}\}_{j=1}^{J},\{\bb{\mu}_{j}\}_{j=1}^{J})$ has the explicit form as:

\begin{equation}
    E_{\bb{\lambda}^{v}}(\widetilde{\bb{z}_{it}}|\bb{y}_{it},w_{it})=\begin{bmatrix}
        \bb{M}_{j}^{v}\bb{\Lambda}_{j}^{v'}\bb{\Psi}^{v^{-1}}(\bb{y}_{it}-\bb{\mu}_{j}^{v})\\
        1
        \end{bmatrix},
\end{equation}

\begin{equation}
    E_{\bb{\lambda}^{v}}(\widetilde{\bb{z}_{it}}\widetilde{\bb{z}_{it}}^{'}|\bb{y}_{it},w_{it})=\begin{bmatrix}
       \bb{M}_{j}^{v}+E_{\bb{\lambda}^{v}}(\bb{z}_{it}|\bb{y}_{it},w_{it})E_{\bb{\lambda}^{v}}(\bb{z}_{it}^{'}|\bb{y}_{it},w_{it}) & E_{\bb{\lambda}^{v}}(\bb{z}_{it}|\bb{y}_{it},w_{it})\\
        E_{\bb{\lambda}^{v}}(\bb{z}_{it}^{'}|\bb{y}_{it},w_{it})& 1
        \end{bmatrix}.
\end{equation}

\subsection*{Appendix B.2: Supplement for M-step}\label{M-add}
Within each M-step, since $h(\bb{\pi})$, $ h(\{\bb{B}_{kj}\}_{k,j=1}^{J})$, and $h(\bb{\Psi},\{\bb{\Lambda}_{j}\}_{j=1}^{J},\{\bb{\mu}_{j}\}_{j=1}^{J})$ do not share parameters, we maximize each of them separately. By solving the first derivative of $h(\bb{\pi})$ equal to 0, the parameters related to the initial state distribution are estimated as follows:
\begin{equation}
    \pi_{j}^{new}=\frac{\sum_{i=1}^{N}\gamma^{v}_{ij}(1)}{\sum_{i=1}^{N}\sum_{k=1}^{J}\gamma^{v}_{ik}(1)}.
\end{equation}
Similarly, the parameters used to characterize the conditional distribution of $\bb{y}_{it}$ given $w_{it}$ are estimated by setting the first derivative of $h(\bb{\Psi},\{\bb{\Lambda}_{j}\}_{j=1}^{J},\{\bb{\mu}_{j}\}_{j=1}^{J})$ equal to 0. $\bb{\Lambda}_{j}$ will be updated as follows:
\begin{equation}
    \widetilde{\bb{\Lambda}_{j}}^{new}=\Big\{\sum_{i=1}^{N}\sum_{t=1}^{T_{i}}\gamma^{v}_{ij}(t)\bb{y}_{it}E_{\bb{\lambda}^{v}}(\widetilde{\bb{z}_{it}}|\bb{y}_{it},w_{it})^{'}\Big\}\Big\{\sum_{i=1}^{N}\sum_{t=1}^{T_{i}}\gamma^{v}_{ij}(t)E_{\bb{\lambda}^{v}}(\widetilde{\bb{z}_{it}}\widetilde{\bb{z}_{it}}^{'}|\bb{y}_{it},w_{it})\Big\}^{-1}.
\end{equation}
Simultaneously, we got the updated estimation of $\bb{\Psi}$ as the following. 
\begin{equation}
    \bb{\Psi}^{new}=\frac{1}{\sum_{i=1}^{N}T_{i}}diag\Big\{\sum_{i=1}^{N}\sum_{t=1}^{T_{i}}\sum_{j=1}^{J}\gamma^{v}_{ij}(t)\{\bb{y}_{it}-\widetilde{\bb{\Lambda}_{j}}^{new}E_{\bb{\lambda}^{v}}(\widetilde{\bb{z}_{it}})\}\bb{y}_{it}^{'}\Big\}.
\end{equation}
Here, since we assume $\bb{\Psi}$ a diagonal matrix, we restrict all the off-diagonal entries of the estimator of $\bb{\Psi}$ to be $0$.

\section*{Appendix C: SEMA} \label{Appendix:2}
\begin{algorithm}[ht]
\hspace*{\SpaceReservedForComments}{}%
\begin{minipage}{\dimexpr\linewidth-\SpaceReservedForComments\relax}
    \caption{Stabilized Expectation-Maximization Algorithm (SEMA)}\label{algo}
   \begin{algorithmic}[1]
    \Procedure{SEMA}{$\{\bb{Y}_{i}\},\{\bb{X}_{i}\}, K, J, \delta_{1},\delta_{2},\bb{\lambda}^0$}\tikzmark{right}
    \State $\bb{\lambda}^{v}\gets \bb{\lambda}^0; \Delta^{1} \gets \delta_{1}+1; \Delta^{2} \gets \delta_{2}+1$
    \While{$\Delta^{1}>\delta_{1}$ and $\Delta^{2}>\delta_{2}$} 
      \State compute $\gamma_{ij}^{v}(t),\epsilon_{ikj}^{v}(t) $ 
      \State update $h(\bb{\pi})$, $h(\{\bb{B}_{kj}\})$, and $h(\bb{\Psi},\{\bb{\Lambda}_{j}\},\{\bb{\mu}_{j}\})$ 
      \State update $\{\bb{\mu}_{j}\},\{\bb{\Lambda}_{j}\},\bb{\Psi},\bb{\pi}$ by optimizing $h(\bb{\pi}), h(\bb{\Psi},\{\bb{\Lambda}_{j}\},\{\bb{\mu}_{j}\})$
      \State update $\{\bb{B}_{kj}\}$ based on $h(\{\bb{B}_{kj}\})$ using stabilized NR/FS
      \State $\Delta^{1} \gets |logP(\{\bb{Y}_{i}\}|\bb{\lambda})-logP(\{\bb{Y}_{i}\}|\bb{\lambda}^{v})|$\tikzmark{right}\Comment{absolute difference in log-likelihood}
      \State $\Delta^{2} \gets \frac{\sum_{l}|\bb{\lambda}_{l}-\bb{\lambda}_{l}^{v}|}{r}$\Comment{average absolute difference in all $r$ free parameters}
      \State $\bb{\lambda}^{v}\gets \bb{\lambda}$
    \EndWhile 
    \State \textbf{return} $\bb{\lambda}$ and $\{w_{it}\} = \{argmax_{j}(\gamma_{ij}(t))\}$
  \EndProcedure
\end{algorithmic}%
\AddNote[blue]{4}{5}{E}
\AddNote[red]{6}{7}{M}
\end{minipage}%
\end{algorithm}


\section*{Appendix D: Synthetic Data Generation}\label{Appendix:3}
To resemble the processed AURORA data with irregular measurements (i.e., CT-EHMFM setting), we uniformly sampled $T_i$ from the interval $[50,100]$ for each subject. Then, we randomly selected $T_i$ integers from the $\{1,\cdots,100\}$ without replacement to get the sequences of the occasions ($t$) that collect measurements. The resulting time intervals $\{\delta_{it}\}$ are then calculated accordingly. To closely replicate the processed AURORA data with only regular measurements (i.e., DT-EHMFM setting), $T_i$ is set to $10$ for all subjects.

With the number of states $J$ and the number of factors $K$ both fixed at three, we first generated data related to the latent states. The initial state of each individual is independently sampled from a multinomial distribution with probability $\bb{\pi}$ = $(\frac{1}{3}, \frac{1}{3}, \frac{1}{3})$. With the initial states, each individual's latent state trajectories are then sampled according to the transition probabilities $\bb{P}_{it}$ or $\bb{P}_{it}(\delta_{it})$ with $\{\bb{B}_{kj}\}$ and ${\bb{x}_{it}^{T}=(x_{it1},x_{it2},x_{it3})}$. Mimicking the AURORA data, we assume all three covariates in the transition model are baseline features that are static over time. While $x_{it1} = 1$ is an intercept, $x_{it2}$ is a binary variable uniformly and independently sampled from $\{0,1\}$ and $x_{it3}$ follows a uniform distribution between $0$ and $1$. Given the dynamic trajectories of states for each individual, suppose that individual $i$ is in the state $j$ at time $t$, the observation vector $\bb{y}_{it}$ is randomly drawn from a normal distribution with a mean of $\bb{\mu}_{j}$ and covariance $\bb{\Lambda}_{j}\bb{\Lambda}_{j}^{'}+\bb{\Psi}$, where $\bb{\Psi} = \bb{I}$. The true values of the unknown parameters are summarized in Table \ref{tab:True}.  

\begin{table}[hbt!]
  \centering
  \resizebox{0.75\textwidth}{!}{\begin{minipage}{\textwidth}
  \centering
\caption{True values of $\bb{\mu}$, $\bb{\Lambda}$, and $\bb{B}_{kj}$ used in simulation studies.}
    \label{tab:True}
\begin{tabular}{cccc} \toprule
    {$j$} & {$1$} & {$2$} & {$3$} \\ \midrule
    $\bb{\mu}_{j}$  & $\begin{pmatrix} 15*\bb{1}_{2} \\ 10*\bb{1}_{3} \\ 5*\bb{1}_{6} \\ 0*\bb{1}_{12}\end{pmatrix}$ & $\begin{pmatrix} 17*\bb{1}_{2} \\ 12*\bb{1}_{3} \\ 7*\bb{1}_{6} \\ 2*\bb{1}_{12}\end{pmatrix}$ & $\begin{pmatrix} 19*\bb{1}_{2} \\ 14*\bb{1}_{3} \\ 9*\bb{1}_{6} \\ 4*\bb{1}_{12}\end{pmatrix}$\\\midrule
    $\bb{\Lambda}_{j}$  & $\begin{pmatrix} 1*\bb{1}_{2} & 1*\bb{1}_{2} &1*\bb{1}_{2} \\ .7*\bb{1}_{5} & 0*\bb{1}_{5} &0*\bb{1}_{5} \\ 0*\bb{1}_{8} & .7*\bb{1}_{8} & 0*\bb{1}_{8} \\ 0*\bb{1}_{8} & 0*\bb{1}_{8} & .7*\bb{1}_{8} \end{pmatrix}$  & $\begin{pmatrix} 0*\bb{1}_{2} & .7*\bb{1}_{2} &0*\bb{1}_{2} \\ 1*\bb{1}_{2} & 1*\bb{1}_{2} & 1*\bb{1}_{2} \\ 0*\bb{1}_{3} & .7*\bb{1}_{3} & 0*\bb{1}_{3} \\ .7*\bb{1}_{8} & 0*\bb{1}_{8} & 0*\bb{1}_{8} \\ 0*\bb{1}_{8} & 0*\bb{1}_{8} & .7*\bb{1}_{8}\end{pmatrix}$ & $\begin{pmatrix} 0*\bb{1}_{4} & 0*\bb{1}_{4} & .7*\bb{1}_{4} \\ 1*\bb{1}_{2} & 1*\bb{1}_{2} &1*\bb{1}_{2} \\ 0 & 0 & .7 \\ 0*\bb{1}_{8} & .7*\bb{1}_{8} & 0*\bb{1}_{8} \\ .7*\bb{1}_{8} & 0*\bb{1}_{8} & 0*\bb{1}_{8}\end{pmatrix}$\\\midrule
    $\bb{B}_{1j}^{T}$ (DT)  & /  & $\begin{pmatrix} -2.95 & -1 & .5 \end{pmatrix}$ & $\begin{pmatrix} -2.95 & -.5 & .5 \end{pmatrix}$\\
    $\bb{B}_{2j}^{T}$ (DT) & $\begin{pmatrix} -2.95 & -.5 & .5\end{pmatrix}$  & / & $\begin{pmatrix} -2.95 & -.5 & .5\end{pmatrix}$\\ 
    $\bb{B}_{3j}^{T}$ (DT) & $\begin{pmatrix} -2.95 & -.5 & .5 \end{pmatrix}$   & $\begin{pmatrix} -2.95 & 1 & .5\end{pmatrix}$ & /\\ \midrule
    $\bb{B}_{1j}^{T}$ (CT)  & /  & $\begin{pmatrix} -2.5 & 1 & -1\end{pmatrix}$ & $\begin{pmatrix} -2.5 & 1 & -1 \end{pmatrix}$\\
    $\bb{B}_{2j}^{T}$ (CT) & $\begin{pmatrix} -3 & 1 & -1 \end{pmatrix}$  & / & $\begin{pmatrix} -2.5 & 1 & -1 \end{pmatrix}$\\ 
    $\bb{B}_{3j}^{T}$ (CT) & $\begin{pmatrix} -3 & 1 & -1\end{pmatrix}$   & $\begin{pmatrix} -3 & 1 & -1 \end{pmatrix}$ & /\\
    \bottomrule
    \end{tabular}
      \end{minipage}}
\end{table}

\section*{Appendix E: Additional Simulation Results}\label{Appendix:4}
\subsection*{Appendix E.1: Estimation Performance of Parameters in TM}
\begin{table}[h]
\centering
\caption{Bias (standard error) of the parameter estimates for each transition model parameter $B_{jkl}$.\label{tab:PPE-TM}}
\begin{tabular}{cccc|ccc}
\toprule
 &  \multicolumn{3}{c}{DT-EHMFM} & \multicolumn{3}{c}{CT-EHMFM}\\
\midrule
$\bb{B}_{jk}$ & $B_{jk0}$ & $B_{jk1}$ & $B_{jk2}$ & $B_{jk0}$ & $B_{jk1}$ & $B_{jk2}$  \\
$\bb{B}_{12}$ & -.050(.495) & -.067(.475) & -.006(.845) & .014(.142) & .017(.150) & -.080(.219) \\
$\bb{B}_{13}$ & -.024(.460) & .032(.408) &	-.065(.700) & -.021(.126) & .010(.125) & .016(.182)\\
$\bb{B}_{21}$ & -.113(.483) & -.028(.409) & .104(.686) &  -.018(.147) & -.004(.139) & .014(.214)\\
$\bb{B}_{23}$ & -.016(.453) & -.020(.426) & .050(.719) & -.003(.112) & .004(.097) & .005(.200)\\
$\bb{B}_{31}$ & -.037(.404) & -.055(.478) & -.007(.614) &  .003(.112) & .002(.107) & -.006(.192)\\
$\bb{B}_{32}$ & -.013(.372) & -.016(.282) & .003(.561) & .006(.131) & -.032(.091) & .028(.232)\\
\bottomrule
\end{tabular}
\end{table}

\subsection*{Appendix E.2: Additional Results for Simulation 1 Under Various Settings}

This section presents additional simulation results that investigate the impact of various factors on estimation performance. Using the simulation settings provided in Appendix D as the baseline setup, we conducted eight additional sets of simulation studies by systematically varying various components. Specifically, for each test, we maintain the baseline setup except for the component under examination. These components include: i) sample size (N), ii) number of measurements for each individual ($T_i$), iii) J, iv) K, v) size of common variance $\bb{\Psi}$, vi) state-to-state difference in $\bb{\mu}_{j}$, vii) state-to-state difference in $\bb{\Lambda}_{j}$, and viii) transition frequency. Following are descriptions of the parameter specifications under various circumstances, followed by the results.


When evaluating the effect of common variance $\bb{\Psi}$, we assess scenarios with the common variance equals .1$\bb{I}$, .5$\bb{I}$, and 1$\bb{I}$ (baseline), respectively. Regarding the effect of $\bb{\mu}$ distinction, we consider two additional settings of $\bb{\mu}$ by adjusting the state-to-state differences in $\bb{\mu}_j$ as specified in Table \ref{mu_diff}.

\begin{table}[H]
    \centering
    \caption{True values of $\bb{\mu}$ with different levels of state-to-state difference} \label{mu_diff}
  \resizebox{.75\textwidth}{!}{\begin{minipage}{\textwidth}
  \centering
\begin{tabular}{cccc} \toprule
    {$j$} & {$1$} & {$2$} & {$3$} \\ \midrule
    $\bb{\mu}$:large diff (baseline)  & $\begin{pmatrix} 15*\bb{1}_{2} \\ 10*\bb{1}_{3} \\ 5*\bb{1}_{6} \\ 0*\bb{1}_{12}\end{pmatrix}$ & $\begin{pmatrix} 17*\bb{1}_{2} \\ 12*\bb{1}_{3} \\ 7*\bb{1}_{6} \\ 2*\bb{1}_{12}\end{pmatrix}$ & $\begin{pmatrix} 19*\bb{1}_{2} \\ 14*\bb{1}_{3} \\ 9*\bb{1}_{6} \\ 4*\bb{1}_{12}\end{pmatrix}$\\\midrule
    $\bb{\mu}$:medium diff  & $\begin{pmatrix} 15*\bb{1}_{2} \\ 10*\bb{1}_{3} \\ 5*\bb{1}_{6} \\ 0*\bb{1}_{12}\end{pmatrix}$ & $\begin{pmatrix} 17*\bb{1}_{2} \\ 12*\bb{1}_{3} \\ 7*\bb{1}_{6} \\ 0*\bb{1}_{12}\end{pmatrix}$ & $\begin{pmatrix} 19*\bb{1}_{2} \\ 14*\bb{1}_{3} \\ 9*\bb{1}_{6} \\ 0*\bb{1}_{12}\end{pmatrix}$\\\midrule
   $\bb{\mu}$:minor diff  & $\begin{pmatrix} 15*\bb{1}_{2} \\ 10*\bb{1}_{3} \\ 5*\bb{1}_{6} \\ 0*\bb{1}_{12}\end{pmatrix}$ & $\begin{pmatrix} 15.75*\bb{1}_{2} \\ 10.75*\bb{1}_{3} \\ 5*\bb{1}_{6} \\ 0*\bb{1}_{12}\end{pmatrix}$ & $\begin{pmatrix} 16.5*\bb{1}_{2} \\ 11.5*\bb{1}_{3} \\ 5*\bb{1}_{6} \\ 0*\bb{1}_{12}\end{pmatrix}$\\
    \bottomrule
    \end{tabular}
     \end{minipage}}
\end{table}

When investigating the effect of $\bb{\Lambda}$ distinction, we consider two additional settings of $\bb{\Lambda}$ by adjusting the state-to-state differences in $\bb{\Lambda}_j$ as outlined in Table \ref{lambda_diff}.

\begin{table}[H]
  \centering
  \caption{True values of $\bb{\Lambda}$ with different state-to-state difference}\label{lambda_diff}
  \resizebox{.75\textwidth}{!}{\begin{minipage}{\textwidth}
  \centering
\begin{tabular}{cccc} \toprule
{$j$} & {$1$} & {$2$} & {$3$} \\ \midrule
    $\bb{\Lambda}$: large diff (baseline)  & $\begin{pmatrix} 1*\bb{1}_{2} & 1*\bb{1}_{2} &1*\bb{1}_{2} \\ .7*\bb{1}_{5} & 0*\bb{1}_{5} &0*\bb{1}_{5} \\ 0*\bb{1}_{8} & .7*\bb{1}_{8} & 0*\bb{1}_{8} \\ 0*\bb{1}_{8} & 0*\bb{1}_{8} & .7*\bb{1}_{8} \end{pmatrix}$  & $\begin{pmatrix} 0*\bb{1}_{2} & .7*\bb{1}_{2} &0*\bb{1}_{2} \\ 1*\bb{1}_{2} & 1*\bb{1}_{2} & 1*\bb{1}_{2} \\ 0*\bb{1}_{3} & .7*\bb{1}_{3} & 0*\bb{1}_{3} \\ .7*\bb{1}_{8} & 0*\bb{1}_{8} & 0*\bb{1}_{8} \\ 0*\bb{1}_{8} & 0*\bb{1}_{8} & .7*\bb{1}_{8}\end{pmatrix}$ & $\begin{pmatrix} 0*\bb{1}_{4} & 0*\bb{1}_{4} & .7*\bb{1}_{4} \\ 1*\bb{1}_{2} & 1*\bb{1}_{2} &1*\bb{1}_{2} \\ 0 & 0 & .7 \\ 0*\bb{1}_{8} & .7*\bb{1}_{8} & 0*\bb{1}_{8} \\ .7*\bb{1}_{8} & 0*\bb{1}_{8} & 0*\bb{1}_{8}\end{pmatrix}$\\\midrule
    $\bb{\Lambda}$: medium diff  & $\begin{pmatrix} .3 & .3 &.3 \\ .7*\bb{1}_{6} & 0*\bb{1}_{6} &0*\bb{1}_{6} \\ 0*\bb{1}_{8} & .7*\bb{1}_{8} & 0*\bb{1}_{8} \\ 0*\bb{1}_{8} & 0*\bb{1}_{8} & .7*\bb{1}_{8} \end{pmatrix}$  & $\begin{pmatrix} .7*\bb{1}_{3} & 0*\bb{1}_{3} &0*\bb{1}_{3} \\ .3 & .3 & .3 \\ .7*\bb{1}_{3} & 0*\bb{1}_{3} & 0*\bb{1}_{3} \\ 0*\bb{1}_{8} & .7*\bb{1}_{8} & 0*\bb{1}_{8} \\ 0*\bb{1}_{8} & 0*\bb{1}_{8} & .7*\bb{1}_{8}\end{pmatrix}$ & $\begin{pmatrix} .7*\bb{1}_{5} & 0*\bb{1}_{5} & 0*\bb{1}_{5} \\ .3 & .3 &.3\\ .7 & 0 & 0 \\ 0*\bb{1}_{8} & .7*\bb{1}_{8} & 0*\bb{1}_{8} \\ 0*\bb{1}_{8} & 0*\bb{1}_{8} & .7*\bb{1}_{8}\end{pmatrix}$\\\midrule
    $\bb{\Lambda}$: minor diff  & $\begin{pmatrix} .05 & .05 &.05 \\ .7*\bb{1}_{6} & 0*\bb{1}_{6} &0*\bb{1}_{6} \\ 0*\bb{1}_{8} & .7*\bb{1}_{8} & 0*\bb{1}_{8} \\ 0*\bb{1}_{8} & 0*\bb{1}_{8} & .7*\bb{1}_{8} \end{pmatrix}$  & $\begin{pmatrix} .7*\bb{1}_{3} & 0*\bb{1}_{3} &0*\bb{1}_{3} \\ .05 & .05 & .05 \\ .7*\bb{1}_{3} & 0*\bb{1}_{3} & 0*\bb{1}_{3} \\ 0*\bb{1}_{8} & .7*\bb{1}_{8} & 0*\bb{1}_{8} \\ 0*\bb{1}_{8} & 0*\bb{1}_{8} & .7*\bb{1}_{8}\end{pmatrix}$ & $\begin{pmatrix} .7*\bb{1}_{5} & 0*\bb{1}_{5} & 0*\bb{1}_{5} \\ .05 & .05 &.05\\ .7 & 0 & 0 \\ 0*\bb{1}_{8} & .7*\bb{1}_{8} & 0*\bb{1}_{8} \\ 0*\bb{1}_{8} & 0*\bb{1}_{8} & .7*\bb{1}_{8}\end{pmatrix}$\\
    \bottomrule
    \end{tabular}
      \end{minipage}}
\end{table}

For the test evaluating the effect of transition frequency, a frequent transition is defined as the probability of remaining in the same state being less than 0.70. The $\bb{B}$ corresponding to the frequent transition is specified in Table \ref{B_diff}, while the $\bb{B}$ in the baseline setting corresponds to infrequent transition.

\begin{table}[H]
  \centering
  \caption{True values of $\bb{B}_{kj}$ with Frequent Transition}\label{B_diff}
  \resizebox{.75\textwidth}{!}{\begin{minipage}{\textwidth}
  \centering
\begin{tabular}{cccc} \toprule
{$j$} & {$1$} & {$2$} & {$3$} \\ \midrule
    $\bb{B}_{1j}^{T}$ (DT): freq transit  & /  & $\begin{pmatrix} -.1.5 & -2 & .75 \end{pmatrix}$ & $\begin{pmatrix} -1.5 & .75 & .5 \end{pmatrix}$\\
    $\bb{B}_{2j}^{T}$ (DT) : freq transit & $\begin{pmatrix} -1.5 & .75 & 1.25\end{pmatrix}$  & / & $\begin{pmatrix} -1.5 & -1.5 & .75\end{pmatrix}$\\ 
    $\bb{B}_{3j}^{T}$ (DT) : freq transit & $\begin{pmatrix} -1.5 & .75 & .75 \end{pmatrix}$   & $\begin{pmatrix} -1.5 & -2 & .75\end{pmatrix}$ & /\\ \midrule
    $\bb{B}_{1j}^{T}$ (CT) : freq transit  & /  & $\begin{pmatrix} .5 & 1 & -.5\end{pmatrix}$ & $\begin{pmatrix} .5 & 1 & -.5 \end{pmatrix}$\\
    $\bb{B}_{2j}^{T}$ (CT) : freq transit& $\begin{pmatrix} -1 & .5 & 1 \end{pmatrix}$  & / & $\begin{pmatrix} -.25 & 1 & -.5 \end{pmatrix}$\\ 
    $\bb{B}_{3j}^{T}$ (CT) : freq transit & $\begin{pmatrix} .5 & .5 & 1\end{pmatrix}$   & $\begin{pmatrix} -.5 & .5 & 1 \end{pmatrix}$ & /\\
    \bottomrule
    \end{tabular}
      \end{minipage}}
\end{table}

For tests evaluating the effects of J, K, N, and $T_i$, the baseline setups are modified as indicated in the following summary tables. Table \ref{tab:DTP} presents the results under the DT settings, while Table \ref{tab:CTP} displays the results under the CT settings.

\begin{table}[H]
  \centering
  \caption{The Mean (standard error) AADs of $\bb{\pi}$, $\bb{\mu}$, $\bb{\Lambda}$, $\bb{\Psi}$, and $\bb{B}$, and $C_{mis}$ of estimations under different DT settings. \label{tab:DTP}}
  \resizebox{.85\textwidth}{!}{\begin{minipage}{\textwidth}
  \centering
\begin{tabular}{lrrrrrr}
\toprule
ADD & $\bb{\pi}$ & $\bb{\mu}$ & $\bb{\Lambda}$ & $\bb{\Psi}$  & $\bb{B}$ & $C_{mis}$\\\hline
$\bb{\Psi} = 1*\bb{I}$ & .027(.014) & .040(.005) & .037(.002) & .030(.005) & .408 (.099) & .002(.001) \\
$\bb{\Psi} = .5*\bb{I}$ & .027(.014) & .033(.005) & .027(.002) & .014(.002) & .407(.095) & .0003(.0004)\\
$\bb{\Psi} = .1*\bb{I}$ & .027(.014) & .026(.006) & .018(.002) & .003(.001) & .407(.095) & .0000(.0000)\\\hline
$\bb{\mu}$: large diff  & .027(.014) & .040(.005) & .037(.002) & .030(.005) & .408 (.099) & .002(.001) \\
$\bb{\mu}$: medium diff & .027(.014) & .040(.005) & .038(.002) & .030(.005) & .426(.103) & .007(.002)\\
$\bb{\mu}$: minor diff & .034(.022) & .089(.056) & .098(.048) & .032(.008) & .714(.338)  & .336(.290)\\\hline
$\bb{\Lambda}$: large diff  & .027(.014) & .040(.005) & .037(.002) & .030(.005) & .408 (.099) & .002(.001) \\
$\bb{\Lambda}$: medium diff & .027(.014) & .038(.004) & .037(.002) & .029(.005) & .422(.101) & .005(.002)\\
$\bb{\Lambda}$: minor diff & .027(.014) & .038(.004) & .046(.002) & .030(.005) & .414(.098)  & .004(.001)\\\hline
$\bb{B}$: infreq transit  & .027(.014) & .040(.005) & .037(.002) & .030(.005) & .408 (.099) & .002(.001) \\
$\bb{B}$: freq transit & .027(.014) & .041(.005) & .039(.002) & .031(.005) & .286\footnote{Note that the ADD of the corresponding estimated transition probability matrix with frequent transit is .033(.012), while that with infrequent transit is .022(.009). Since for each individual, there are only 10 observations, frequent transition will definitely help estimating the transition model with more different observations. However, the estimated probability matrix might be affected differently.}(.080) & .005(.002)\\\hline

$J=2$ & .031(.023) & .032(.005) & .031(.002) & .029(.005) & .340 (.132) & .001(.001) \\
$J=3$ & .027(.014) & .040(.005) & .037(.002) & .030(.005) & .408 (.099) & .002(.001) \\
$J=4$& .025(.010) & .046(.004) & .043(.002) & .030(.004) & .512(.094) & .003(.001)\\\hline

$K=2$ & .027(.014) & .040(.006) & .037(.002) & .029(.005) & .421 (.098) & .006(.002) \\
$K=3$& .027(.014) & .040(.005) & .037(.002) & .030(.005) & .408 (.099) & .002(.001) \\
$K=5$& .027(.014) & .041(.005) & .040(.002) & .034(.005) & .408(.095) & .0004(.0004)\\\hline
$N = 50; T= 10$ & .057(.027) & .081(.012) & .077(.005) & .061(.010) & 1.199 (.496) & .0029(.0022) \\
$N = 100; T = 10$ & .038(.020) & .057(.008) & .053(.003) & .043(.006) & .648 (.226) & .0025(.0014) \\
$N = 200; T = 10$& .027(.014) & .040(.005) & .037(.002) & .030(.005) & .408 (.099) & .0023(.0010) \\
$N = 300; T = 10$ & .022(.011) & .033(.004) & .031(.002) & .024(.004) & .340 (.083) & .0023(.0009) \\
$N = 700; T = 10$& .014(.008) & .021(.003) & .020(.001) & .016(.003) & .206(.043) & .0021(.0005)\\\hline
$N = 200; T = 3$& .027(.014) & .074(.009) & .069(.004) & .056(.008) & 1.422 (.607) & .0044(.0029) \\
$N = 200; T = 5$& .027(.014) & .058(.007) & .053(.003) & .043(.006) & .707 (.261) & .0028(.0017) \\
$N = 200; T = 10$& .027(.014) & .040(.005) & .037(.002) & .030(.005) & .408 (.099) & .0023(.0010) \\
$N = 200; T = 20$& .026(.014) & .029(.004) & .027(.002) & .021(.003) & .280 (.073) & .0018(.0007) \\
$N = 200; T = 200$& .027(.014) & .009(.001) & .009(.0005) & .006(.001) & .082(.020) & .0017(.0002)\\
\bottomrule
\end{tabular}
      \end{minipage}}
\end{table}


\begin{table}[H]
  \centering
  \caption{The Mean (standard error) AADs of $\bb{\pi}$, $\bb{\mu}$, $\bb{\Lambda}$, $\bb{\Psi}$, and $\bb{B}$, and $C_{mis}$ of estimations under different CT settings. \label{tab:CTP}}
  \resizebox{.85\textwidth}{!}{\begin{minipage}{\textwidth}
  \centering
\begin{tabular}{lrrrrrr}
\toprule
ADD & $\bb{\pi}$ & $\bb{\mu}$ & $\bb{\Lambda}$ & $\bb{\Psi}$  & $\bb{B}$ & $C_{mis}$\\\hline
$\bb{\Psi} = 1*\bb{I}$ & .026(.013) & .015(.002) & .014(.001) & .011(.002) & .120(.027) & .0024(.0005) \\
$\bb{\Psi} = .5*\bb{I}$ & .026(.013) & .012(.002) & .011(.001) & .005(.001) & .119(.027) & .0003(.0001)\\
$\bb{\Psi} = .1*\bb{I}$ & .026(.013) & .010(.003) & .007(.001) & .001(.000) & .119(.027) & .0000(.0000)\\\hline

$\bb{\mu}$: large diff  & .026(.013) & .015(.002) & .014(.001) & .011(.002) & .120(.027) & .002(.0005) \\
$\bb{\mu}$: medium diff & .027(.013) & .015(.001) & .015(.001) & .011(.002) & .124(.028) & .007(.001)\\
$\bb{\mu}$: minor diff & .034(.017) & .017(.003) & .478(.057) & .011(.002) & .161(.040)  & .085(.004)\\\hline

$\bb{\Lambda}$: large diff  & .026(.013) & .015(.002) & .014(.001) & .011(.002) & .120(.027) & .002(.0005) \\
$\bb{\Lambda}$: medium diff & .026(.013) & .014(.002) & .014(.001) & .011(.002) & .122(.029) & .006(.001)\\
$\bb{\Lambda}$: minor diff & .026(.013) & .014(.002) & .014(.001) & .011(.002) & .122(.028)  & .004(.001)\\\hline

$\bb{B}$: infreq transit  & .026(.013) & .015(.002) & .014(.001) & .011(.002) & .120(.027) & .002(.0005) \\
$\bb{B}$: freq transit & .027(.013) & .015(.002) & .014(.001) & .011(.002) & .137(.043) & .007(.001)\\\hline

$J=2$ & .027(.021) & .012(.002) & .012(.001) & .011(.002) & .087(.026) & .0015(.0003) \\
$J=3$ & .026(.013) & .015(.002) & .014(.001) & .011(.002) & .120(.027) & .0024(.0005) \\
$J=4$& .024(.011) & .017(.002) & .016(.001) & .011(.001) & .150(.028) & .0032(.0005)\\\hline

$K=2$ & .026(.013) & .015(.002) & .014(.001) & .010(.002) & .124(.027) & .0069(.0008) \\
$K=3$& .026(.013) & .015(.002) & .014(.001) & .011(.002) & .120(.027) & .0024(.0005) \\
$K=5$& .026(.013) & .015(.002) & .015(.001) & .012(.002) & .119(.027) & .0005(.0002)\\\hline
$N = 50; T_{i}\in[50,100]$ & .052(.029) & .030(.004) & .028(.002) & .022(.003) & .254(.058) & .0027(.0008) \\
$N = 100; T_{i}\in[50,100]$ & .042(.019) & .021(.003) & .020(.001) & .015(.002) & .175(.038) & .0025(.0006)  \\
$N = 200; T_{i}\in[50,100]$& .026(.013) & .015(.002) & .014(.001) & .011(.002) & .120(.027) & .0024(.0005) \\
$N = 500; T_{i}\in[50,100]$ & .018(.010) & .009(.001) & .009(.001) & .007(.001) & .076(.017) & .0024(.0003) \\
\hline
$N = 200; T_{i}\in[10,30]$& .029(.014) & .029(.004) & .027(.002) & .021(.004) & .251(.061) & .0028(.0009) \\
$N = 200; T_{i}\in[30,50]$& .027(.014) & .020(.003) & .019(.001) & .015(.003) & .180(.043) & .0025(.0005) \\
$N = 200; T_{i}\in[50,100]$& .026(.013) & .015(.002) & .014(.001) & .011(.002) & .120(.027) & .0024(.0005) \\
$N = 200; T_{i}\in[100,150]$& .028(.015) & .012(.001) & .011(.001) & .008(.001) & .098(.021) & .0023(.0003) \\
\bottomrule
\end{tabular}
      \end{minipage}}
\end{table}



\subsection*{Appendix E.3: Additional Results for Simulation 3}
This section investigates the performance of BIC/AIC in model selection for DT-EHMFM with varying sample sizes. We employ the same data generation process as in Simulation 3 discussed in the main paper, with the only difference being the number of observations per individual. Table \ref{fig:sample size and BICAIC} summarizes the percentage of instances where a model with accurate $J$ and $K$ was selected. While BIC consistently recommended the model with accurate $J$ and $K$, AIC increased the likelihood of recommending a model with accurate $J$ and $K$ as the sample size increased.
\begin{table}[H]
\caption{Percentage of correct model selection of $(J,K)$ based on BIC/AIC} \label{fig:sample size and BICAIC}
\begin{center}
\begin{tabular}{rrrrrr}
(N,T) & AIC & BIC \\\hline
(200,10) & 94\% & 100\%\\
(200,40) & 99\% & 100\%\\
(200,200) & 100\% & 100\%\\\hline
\end{tabular}
\end{center}
\end{table}

\section*{Appendix F: Additional Real Analysis Results}\label{Appendix:5}
This section presents all of the parameter estimates for the model with $J=3$ and $K=8$ that was fitted. Included in Table \ref{fitted mu} is the estimated $\bb{\mu}_j$ for each state. The estimated loading matrix for state 1 is provided in Table \ref{fitted load1}, while that for state 2 and state 3 are provided in Table \ref{fitted load2} and Table \ref{fitted load3}, respectively. Finally, the parameter estimates for the transition model are displayed in Table \ref{fitted B}.

\begin{table}[H]
  \centering
  \resizebox{.9\textwidth}{!}{\begin{minipage}{\textwidth}
  \centering
        \caption{Estimated $\bb{\mu}_{j}$}\label{fitted mu}
\begin{tabular}{lrrr}
\toprule
$\bb{\mu}_{j}$ &         j=1 &         j=2 &         j=3 \\
\midrule
meanAcc                    &     7.146 &     7.231  &     6.860 \\
amplitude                  &     3.906 &     3.913  &     3.621\\
SWCK                       &     3.269 &     3.220  &     3.151 \\
L5                          &     3.765 &     3.800 &     3.555 \\
NNmean.min                  &     7.663 &     7.496  &     6.881\\
NNmean.mean                 &     8.406 &     7.923 &     6.960\\
NNskew.q3                   &     0.998 &     0.956 &     1.305\\
SDNN.max                    &     7.284 &     6.571  &     6.246\\
SDNN.min                    &    28.705 &    27.628 &    26.944\\
SDNN.mean                   &    12.673 &    11.584 &    10.616\\
SDNN.var                    &    20.014 &    19.333 &    18.512\\
lfhf.min                    &    -4.393 &    -4.428 &    -4.829 \\
lfhf.mean                   &    -0.125 &    -0.046 &    -0.642\\
dc.min                      &     3.888 &     3.168 &     2.706\\
dc.mean                     &     7.423 &     6.453 &     5.418\\
SD1SD2.max                  &   101.371 &   101.620 &   102.439\\
SD1SD2.min                  &    26.001 &    25.885 &    26.745\\
SD1SD2.mean                 & 1,703.646 & 1,703.711 & 1,704.842\\
SD1SD2.var                  &    23.038 &    23.197 &    23.943\\
ApEn.max                    &     4.320 &     3.813 &     3.975\\
ApEn.min                    &    -2.694 &    -1.915 &    -2.280\\
ApEn.mean                   &     1.901 &     1.596 &     1.146\\
ApEn.var                    &    -4.546 &    -5.396 &    -4.782\\
\bottomrule
\end{tabular}
      \end{minipage}}
\end{table}

\begin{table}[H]
  \centering
  \caption{Estimated Standardized Promax Factor Loading Matrix: State = 1 ($\bb{\Lambda_{1}}$)}\label{fitted load1}
  \resizebox{.75\textwidth}{!}{\begin{minipage}{\textwidth}
  \centering
\begin{tabular}{lrrrrrrrr}
\toprule
{} &      0 &      1 &      2 &      3 &      4 &      5 &      6 &      7 \\
\midrule
meanAcc     & -0.014 &  0.022 & -0.013 & -0.032 & \bftab{-0.989} & -0.049 &  0.032 &  0.001 \\
amplitude   &  0.011 &  0.023 &  0.004 &  0.046 & \bftab{-0.808} & -0.017 & -0.090 & -0.066 \\
SWCK        &  0.020 &  0.035 &  0.013 & -0.168 &  \bftab{0.435} &  0.039 &  0.214 &  0.057 \\
L5          &  0.008 &  0.009 &  0.009 & -0.126 & \bftab{-0.445} &  0.017 &  0.183 &  0.067 \\
NNmean.min  &  0.002 &  0.109 & -0.085 &  \bftab{0.860} &  0.011 & -0.065 &  0.189 &  0.097 \\
NNmean.mean &  0.187 &  0.278 &  0.100 &  \bftab{0.752} & -0.025 &  0.052 &  0.215 &  0.013 \\
NNskew.q3   &  0.066 &  0.043 & -0.122 & -0.388 & -0.009 & -0.379 & -0.354 &  0.062 \\
SDNN.max    &  \bftab{0.519} & -0.007 &  0.028 &  0.097 &  0.053 &  0.267 & -0.053 & -0.057 \\
SDNN.min    & -0.052 & -0.182 & -0.133 &  0.280 &  0.004 & -0.181 &  \bftab{0.918} & -0.252 \\
SDNN.mean   &  \bftab{.517} &  0.255 & -0.068 & -0.047 & -0.004 & -0.093 &  \bftab{0.470} & -0.162 \\
SDNN.var    &  \bftab{0.672} &  0.069 & -0.010 & -0.052 & -0.069 &  \bftab{0.469} & -0.268 & -0.036 \\
lfhf.min    &  0.112 & -0.115 & \bftab{-1.030} &  0.104 &  0.000 & -0.048 &  0.175 &  0.217 \\
lfhf.mean   &  0.000 &  0.017 & \bftab{-0.973} & -0.046 & -0.019 &  0.349 &  0.046 &  0.255 \\
dc.min      & -0.039 & -0.146 & -0.004 &  0.038 & -0.038 &  0.185 &  0.164 & \bftab{-0.719} \\
dc.mean     &  0.133 &  0.104 &  0.155 & -0.148 &  0.015 & -0.027 &  0.181 & \bftab{-0.929} \\
SD1SD2.max  & -0.014 & -0.029 &  \bftab{0.514} &  0.081 &  0.039 & -0.062 &  0.045 &  \bftab{0.489} \\
SD1SD2.min  &  0.036 &  0.006 &  0.184 & -0.003 & -0.041 & \bftab{-0.787} &  0.150 &  0.088 \\
SD1SD2.mean &  0.095 & -0.121 &  \bftab{0.727} &  0.074 & -0.002 & \bftab{-0.417} &  0.041 &  0.191 \\
SD1SD2.var  &  0.058 & -0.128 &  \bftab{0.655} &  0.014 & -0.050 &  0.346 & -0.047 &  0.345 \\
ApEn.max    &  0.106 & \bftab{-0.737} &  0.005 & -0.334 &  0.019 & -0.047 &  0.204 & -0.007 \\
ApEn.min    & \bftab{-0.991} &  0.024 &  0.012 & -0.086 & -0.030 &  0.163 &  0.103 & -0.020 \\
ApEn.mean   & \bftab{-0.472} & \bftab{-0.696} &  0.038 & -0.160 &  0.011 &  0.117 &  0.053 & -0.116 \\
ApEn.var    &  \bftab{1.063} & -0.246 & -0.024 & -0.090 & -0.030 & -0.119 &  0.117 & -0.017 \\
\bottomrule
\end{tabular}
      \end{minipage}}
\end{table}

\begin{table}[H]
  \centering
  \caption{Estimated Standardized Promax Factor Loading Matrix: State = 2 ($\bb{\Lambda_{2}}$)}\label{fitted load2}
  \resizebox{.75\textwidth}{!}{\begin{minipage}{\textwidth}
  \centering
\begin{tabular}{lrrrrrrrr}
\toprule
{} &      0 &      1 &      2 &      3 &      4 &      5 &      6 &      7 \\
\midrule
meanAcc     & -0.006 &  0.003 &  0.010 & -0.034 &  \bftab{0.938} & -0.001 &  0.058 & -0.232 \\
amplitude   &  0.027 & -0.034 &  0.005 &  0.038 &  \bftab{0.860} & -0.003 & -0.015 & -0.077 \\
SWCK        & -0.084 &  0.111 &  0.031 & -0.130 & \bftab{-0.672} &  0.015 &  0.226 & -0.084 \\
L5          & -0.062 &  0.079 & -0.013 & -0.141 &  0.128 &  0.042 &  0.142 & -0.268 \\
NNmean.min  & -0.121 &  0.043 &  0.003 &  \bftab{0.807} &  0.047 & -0.043 & -0.000 & -0.035 \\
NNmean.mean & -0.033 & -0.055 & -0.079 &  \bftab{0.807} &  0.069 &  0.016 &  0.318 &  0.102 \\
NNskew.q3   &  0.093 & -0.185 &  0.114 & \bftab{-0.406} & -0.093 & -0.161 & -0.099 & -0.343 \\
SDNN.max    &  0.019 & -0.025 & -0.197 &  0.063 & -0.129 & -0.097 &  \bftab{0.600} & -0.027 \\
SDNN.min    &  \bftab{0.638} &  0.345 &  0.188 &  0.253 & -0.033 &  0.044 &  0.057 & -0.048 \\
SDNN.mean   &  \bftab{0.530} & -0.043 &  0.388 &  0.137 & -0.040 &  0.089 &  \bftab{0.663} &  0.045 \\
SDNN.var    & -0.100 & -0.120 & -0.010 &  0.057 &  0.037 &  0.013 &  \bftab{0.862} & -0.091 \\
lfhf.min    & -0.107 & -0.117 &  \bftab{0.869} &  0.111 & -0.029 & -0.158 & -0.118 & -0.154 \\
lfhf.mean   & -0.135 & -0.029 &  \bftab{0.980} & -0.109 & -0.026 &  0.087 & -0.131 &  0.128 \\
dc.min      &  \bftab{0.473} &  0.148 & -0.011 & -0.082 &  0.110 & -0.202 & -0.047 &  0.069 \\
dc.mean     &  \bftab{0.770} & -0.112 & -0.058 & -0.088 &  0.100 & -0.318 &  0.283 &  0.129 \\
SD1SD2.max  & -0.038 & -0.079 & -0.212 &  0.072 & -0.056 &  \bftab{0.675} & -0.093 &  0.003 \\
SD1SD2.min  &  0.103 & -0.061 & -0.233 &  0.085 & -0.035 & -0.177 & -0.291 & \bftab{-0.416} \\
SD1SD2.mean &  0.042 & -0.044 & \bftab{-0.652} &  0.157 & -0.041 &  0.263 & -0.093 & -0.287 \\
SD1SD2.var  & -0.077 & -0.039 & -0.171 & -0.027 &  0.041 &  \bftab{0.834} &  0.022 &  0.095 \\
ApEn.max    &  \bftab{0.837} & -0.290 & -0.094 & -0.156 & -0.022 &  0.208 & -0.047 & -0.029 \\
ApEn.min    &  \bftab{0.514} &  \bftab{0.487} & -0.024 & -0.070 & -0.026 &  0.059 & -0.250 & -0.091 \\
ApEn.mean   &  \bftab{0.932} &  0.070 & -0.230 & -0.047 & -0.019 &  0.024 & -0.033 & -0.040 \\
ApEn.var    &  0.002 & \bftab{-0.972} &  0.091 & -0.069 &  0.098 &  0.096 &  0.114 &  0.037 \\
\bottomrule
\end{tabular}
      \end{minipage}}
\end{table}

\begin{table}[H]
  \centering
  \caption{Estimated Standardized Promax Factor Loading Matrix: State = 3 ($\bb{\Lambda_{3}}$)}\label{fitted load3}
  \resizebox{.75\textwidth}{!}{\begin{minipage}{\textwidth}
  \centering
\begin{tabular}{lrrrrrrrr}
\toprule
{} &      0 &      1 &      2 &      3 &      4 &      5 &      6 &      7 \\
\midrule
meanAcc     &  0.008 & -0.008 &  0.010 & -0.041 & \bftab{-0.989} &  0.004 &  0.049 &  0.159 \\
amplitude   &  0.010 &  0.001 &  0.090 &  0.007 & \bftab{-0.725} &  0.025 & -0.053 &  0.380 \\
SWCK        & -0.003 & -0.086 &  0.099 & -0.132 &  0.106 &  0.008 &  0.165 & \bftab{-0.560} \\
L5          &  0.009 & -0.040 & -0.065 & -0.058 & \bftab{-0.609} & -0.018 &  0.095 & -0.261 \\
NNmean.min  &  0.026 & -0.034 & -0.078 &  \bftab{0.686} &  0.047 & -0.133 &  0.056 &  0.005 \\
NNmean.mean & -0.092 & -0.041 &  0.007 &  \bftab{0.869} &  0.013 &  0.007 &  0.230 &  0.154 \\
NNskew.q3   & -0.050 & -0.044 & -0.007 & \bftab{-0.533} & -0.018 & -0.061 &  0.327 & -0.218 \\
SDNN.max    &  0.066 &  0.010 & -0.130 & -0.028 &  0.024 & -0.096 &  \bftab{0.668} & -0.174 \\
SDNN.min    & \bftab{-0.780} & -0.344 &  0.061 &  0.103 & -0.005 &  0.059 & -0.071 & -0.096 \\
SDNN.mean   & \bftab{-0.798} &  0.048 &  0.286 &  0.227 & -0.048 &  0.155 &  0.378 & -0.086 \\
SDNN.var    & -0.069 &  0.120 & -0.019 &  0.134 & -0.105 &  0.033 &  \bftab{0.945} & -0.045 \\
lfhf.min    &  0.049 &  0.082 &  \bftab{0.968} &  0.062 & -0.040 & -0.180 & -0.181 & -0.271 \\
lfhf.mean   &  0.017 &  0.005 &  \bftab{1.073} & -0.108 & -0.007 &  0.040 & -0.139 & -0.069 \\
dc.min      & \bftab{-0.408} & -0.034 & -0.077 & -0.125 &  0.005 & -0.189 & -0.118 &  0.171 \\
dc.mean     & \bftab{-0.750} &  0.049 & -0.099 & -0.007 &  0.028 & -0.171 &  0.274 &  0.223 \\
SD1SD2.max  &  0.035 &  0.026 & -0.282 &  0.011 &  0.007 &  \bftab{0.585} & -0.094 & -0.105 \\
SD1SD2.min  & -0.122 &  0.112 & -0.295 &  0.117 & -0.081 & -0.198 & -0.333 & -0.356 \\
SD1SD2.mean &  0.053 &  0.070 & \bftab{-0.610} &  0.054 & -0.019 &  0.227 & -0.126 & -0.244 \\
SD1SD2.var  & -0.027 & -0.039 & -0.158 & -0.134 & -0.009 &  \bftab{0.925} & -0.042 &  0.095 \\
ApEn.max    & \bftab{-0.837} &  \bftab{0.490} &  0.018 & -0.150 &  0.047 &  0.081 & -0.086 & -0.013 \\
ApEn.min    & \bftab{-0.801} & -0.357 & -0.038 & -0.013 & -0.002 &  0.003 & -0.157 & -0.090 \\
ApEn.mean   & \bftab{-1.009} &  0.041 & -0.079 &  0.029 &  0.003 &  0.077 & -0.021 & -0.117 \\
ApEn.var    &  0.256 &  \bftab{0.971} &  0.021 & -0.027 &  0.028 & -0.024 &  0.127 &  0.152 \\
\bottomrule
\end{tabular}
      \end{minipage}}
\end{table}


\begin{table}[H]
    \centering
    \caption{Estimated Transition Model ($\bb{B}_{kj}$)}\label{fitted B}
  \resizebox{.75\textwidth}{!}{\begin{minipage}{\textwidth}
  \centering
    \begin{tabular}{lrrrrrrrrr}
\toprule
{j} &      $\bb{B}_{1j0}$ &      $\bb{B}_{1j1}$ &      $\bb{B}_{1j2}$ & $\bb{B}_{2j0}$ &      $\bb{B}_{2j1}$ &      $\bb{B}_{2j2}$ & $\bb{B}_{3j0}$ &      $\bb{B}_{3j1}$ &      $\bb{B}_{3j2}$ \\
\midrule
1 &  - &  - &  - & -0.462 & -0.243 & -0.019  & -2.430 & -0.233 & -0.018\\
2 & -1.736 & -0.091 & 0.017 &  - &  - &  -  &  0.334 & -0.759 & -0.021\\
3 & -5.046 &  1.426 & 0.015  & -1.723 &  0.053 &  0.005 &  - &  - &  -\\
\bottomrule
\end{tabular}
    \label{tab:my_label}
    \end{minipage}}
\end{table}

\section*{Appendix G: Flash Survey Data}\label{Appendix:6}

Installed on the participants' smartphones, the Mindstrong DiscoveryTM application was used to deliver brief questionnaires (flash surveys). Participants were asked to respond daily for the first week of the study, then every other day until week 12, after which they were asked to respond weekly. 

Based on the RDoC framework, ten latent constructs associated with APNS were developed using flash survey items selected by domain experts: Pain, Loss, Sleep Discontinuity, Nightmare, Somatic Symptoms, Mental Fatigue, Avoidance, Re-experience, and Anxious. Participant scores for each construct were calculated based on their responses to the flash survey questions, with higher scores indicating more severe symptoms. While most construct scores range from $-1$ to $5$, the score range for mental fatigue and somatic symptoms is $(0,12)$, and the range of the pain construct score is $(-1,10)$. For analytical purposes, we scaled each construct to fall within the range of $[0,1]$. Moreover, we only consider observations for each individual whose estimated states are known on the same day they submitted survey responses.


\end{document}